\documentclass{aa}
\usepackage{graphicx}
\usepackage{color}
\usepackage{natbib}
\usepackage{txfonts}
\begin{document} 
\newcommand{\hagrid}{{\sc HaGrid} } % define how we write hagrid
\newcommand{\Ha}{H$\alpha$ }
\newcommand{\URGO}{U_{\rm RGO}}

   \title{High resolution H$\alpha$ imaging of the Northern Galactic Plane, and the IGAPS images database}
   \titlerunning{H$\alpha$ northern Galactic plane imaging}

   \author{R. Greimel\inst{\ref{i:rg}} \and % contributed
          J. E. Drew\inst{\ref{i:herts},\ref{i:UCL}} \and  % contributed
          M. Mongui\'o\inst{\ref{i:herts},\ref{i:icc},\ref{i:upc}} \and % contributed/commented
          R. P. Ashley\inst{\ref{i:ING}} \and % confirmed/read
          G. Barentsen\inst{\ref{i:herts},\ref{i:bay}} \and % confirmed/commented
          J. Eisl\"offel\inst{\ref{i:TLS}} \and % contributed/commented
          A. Mampaso\inst{\ref{i:IAC}} \and % contributed/commented
          R. A. H. Morris\inst{\ref{i:bristol}} \and %contributed/commented
          T. Naylor\inst{\ref{i:exeter}} \and % confirmed/commented
          C. Roe\inst{\ref{i:herts}} \and % confirmed/read
          L. Sabin\inst{\ref{i:UNAM}} \and % confirmed/commented
          B. Stecklum\inst{\ref{i:TLS}} \and % contributed/commented
          N. J. Wright\inst{\ref{i:keele}} \and % contributed
          P. J. Groot\inst{\ref{i:nijmegen},\ref{i:cape},\ref{i:SAAO},\ref{i:dia_cape}} \and % confirmed/commented
          M. J. Irwin\inst{\ref{i:cambridge}} \and % commented
          M. J. Barlow\inst{\ref{i:UCL}} \and % commented
          C. Fari\~na\inst{\ref{i:ING},\ref{i:IAC}} \and % confirmed/commented
          A. Fern\'andez-Mart\'in\inst{\ref{i:IAC},\ref{i:CAHA}} \and %confirmed/commented
%          J. Casares\inst{11,12} \and
%          B. T. G\"ansicke\inst{13}\and
%P. J. Carter\inst{13,14}\and
%J. M. Corral-Santana\inst{11,15} \and
%N. P. Gentile-Fusillo\inst{13,15}\and
%S. Greiss\inst{13} \and
%L. M. van Haaften\inst{6,16}\and
%M. Hollands\inst{13}\and
%D. Jones\inst{11,12}\and
%T. Kupfer\inst{6,17}\and
%C. J. Manser\inst{13}\and
%D. N. A. Murphy\inst{10}\and
%A. F. McLeod\inst{6,16,18}\and
%T. Oosting\inst{6}\and
%A. Mampaso\inst{13} \and
         Q. A. Parker\inst{\ref{i:hongkong},\ref{i:hongkong_sr}} \and % confirmed
         S. Phillipps\inst{\ref{i:bristol}} \and % confirmed
%S. Pyrzas\inst{13,20}\and
%P. Rodr\'iguez-Gil\inst{11,12}\and
%J. van Roestel\inst{6,21}\and
         S. Scaringi\inst{\ref{i:durham}} \and % confirmed
         A. A. Zijlstra\inst{\ref{i:jodrell}},\inst{\ref{i:hongkong_sr}} % confirmed/commented
%P. Schellart\inst{6}\and
%O. Toloza\inst{13}\and
%O. Vaduvescu\inst{11,22}\and
%L. van Spaandonk\inst{13,23}\and
%K. Verbeek\inst{6}\and
%J. Fabregat\inst{26}\and 
%A. Harris\inst{2}\and
%R. Raddi\inst{13,28}\and
%Y. Unruh\inst{30}\and
%J. S Vink\inst{31}\and
%R. Wesson\inst{4} \and
%A. Cardwell\inst{22,32}\and 
%R. K. Cochrane\inst{22}\and
%S. Doostmohammadi \inst{22,33}\and
%T. Mocnik\inst{22}\and 
%H. Stoev\inst{22}\and
%L. Su\'arez-Andr\'es\inst{22}\and
%V. Tudor \inst{22}\and
%T. G. Wilson\inst{22}\and
%T. J. Zegmott\inst{22}
}

\institute{
{IGAM, Institute of Physics, University of Graz, Universitätsplatz 5/II, 8010 Graz, Austria}\label{i:rg} \and
{School of Physics, Astronomy \& Mathematics, University of Hertfordshire, Hatfield AL10 9AB, UK}\label{i:herts} \and
{Department of Physics \& Astronomy, University College London, Gower Street, London WC1E 6BT, UK}\label{i:UCL} \and
{Institut d'Estudis Espacials de Catalunya, Universitat de Barcelona (ICC-UB), Mart\'i i Franqu\`es 1,
E-08028 Barcelona, Spain}\label{i:icc} \and
{Universitat Polit\`ecnica de Catalunya, Departament de F\'isica, c/Esteve Terrades 5, 08860 Castelldefels, Spain}\label{i:upc} \and
{Isaac Newton Group, Apartado de correos 321, E-38700 Santa Cruz de La Palma, Canary Islands, Spain}\label{i:ING} \and
{Bay Area Environmental Research Institute, P.O. Box 25, Moffett Field, CA 94035, USA}\label{i:bay} \and
{Th\"uringer Landessternwarte, Sternwarte 5, 07778 Tautenburg, Germany}\label{i:TLS} \and
{Instituto de Astrof\'isica de Canarias, E-38205 La Laguna, Tenerife, Spain}\label{i:IAC} \and
{Astrophysics Group, School of Physics, University of Bristol, Tyndall Av, Bristol, BS8 1TL, UK}\label{i:bristol} \and
{School of Physics, University of Exeter, Exeter EX4 4QL, UK}\label{i:exeter} \and
{Instituto de Astronomía, Universidad Nacional Autónoma de México (UNAM), AP 106,  Ensenada 22800, BC, México}\label{i:UNAM} \and
{Astrophysics Group, Keele University, Keele, ST5 5BG, UK}\label{i:keele} \and
{Department of Astrophysics/IMAPP, Radboud University, P.O. Box 9010, 6500 GL Nijmegen, The Netherlands}\label{i:nijmegen} \and
{Department of Astronomy, University of Cape Town, Private Bag X3, Rondebosch, 7701, South Africa}\label{i:cape} \and
{South African Astronomical Observatory, P.O. Box 9, Observatory, 7935, South Africa}\label{i:SAAO} \and
{The Inter-University Institute for Data Intensive Astronomy, University of Cape Town, Private Bag X3, Rondebosch, 7701, South Africa}\label{i:dia_cape} \and
{Institute of Astronomy, University of Cambridge, Madingley Road, Cambridge, CB3 0HA, UK}\label{i:cambridge} \and
{Centro Astron\'omico Hispano-Alem\'an, Observatorio de Calar Alto, Sierra de los Filabres s/n, E-04550 G\'ergal, Almeria, Spain}\label{i:CAHA} \and
{The University of Hong Kong, Department of Physics, Hong Kong SAR, China}\label{i:hongkong} \and
{The Laboratory for Space Research, Cyberport 4, Hong Kong SAR, China}\label{i:hongkong_sr} \and
%Departamento de Astrof\'isica, Universidad de La Laguna, E-38206 La Laguna, Tenerife, Spain\and
%University of Warwick, Department of Physics, Gibbet Hill Road, Coventry, CV4 7AL, UK\and
%Department of Earth and Planetary Sciences, University of California, Davis, One Shields Avenue, Davis, CA 95616, USA\and
%European Southern Observatory (ESO), Av. Alonso de C\'ordova 3107, 7630355 Vitacura, Santiago, Chile\and
%Department of Physics and Astronomy, Texas Tech University, PO Box 41051, Lubbock, TX 79409, USA\and
%Kavli Institute for Theoretical Physics, University of California, Santa Barbara, CA 93106, USA\and
%Department of Astronomy, University of California Berkeley, Berkeley, CA 94720, USA\and
%Hamad Bin Khalifa University (HBKU), Qatar Foundation, P.O. Box 5825, Doha, Qatar.\and
%Division of Physics, Mathematics and Astronomy, California Institute of Technology, Pasadena, CA 91125, USA \and
%Mollerlyceum, 4611DX, Bergen op Zoom, The Netherlands\and
%Observatorio Astron\'omico, Universidad de Valencia, Calle Catedr\'atico Jos\'e Beltr\'an 2, 46980 Paterna, Spain \and
%Dr. Remeis-Sternwarte, Friedrich Alexander Universit\"at Erlangen-N\"urnberg, 
%Sternwartstr 7, D-96049 Bamberg, Germany\and
%Department of Physics, Imperial College London, SW7 2AZ \and
%Armagh Observatory and Planetarium, BT61 9DG, Armagh, UK\and
%LBT Observatory, University of Arizona, 933 N. Cherry Ave, Tucson, AZ 85721-0009, U.S.A. \and
%Department of Physics, Shahid Bahonar University of Kerman, Iran     
{Department of Physics, University of Durham, South Road, Durham, DH1 3LE}\label{i:durham} \and
{Jodrell Bank Centre for Astrophysics, Alan Turing Building, Manchester M13 9PL, UK}\label{i:jodrell}
}

   \date{Received March 31, 2021; accepted July 13, 2021}

  \abstract {The INT Galactic Plane Survey (IGAPS) is the merger of the optical photometric surveys, IPHAS and UVEX, based on data from the Isaac Newton Telescope (INT) obtained between 2003 and 2018. These capture the entire northern Galactic plane within the Galactic coordinate range, $|b| < 5^{\circ}$ and $30^{\circ} < \ell < 215^{\circ}$. From the beginning, the incorporation of narrowband H$\alpha$ imaging has been a unique and distinctive feature of this effort. Alongside a focused discussion of the nature and application of the H$\alpha$ data, we present the IGAPS world-accessible database of images for all 5 survey filters, $i$, $r$, $g$, $\URGO$ and narrowband $H\alpha$, observed on a pixel scale of 0.33 arcsec and at an effective (median) angular resolution of 1.1--1.3 arcsec. %The database of over half a million images of individual CCD frames is available to the world community via a website 
  %incorporating a simple user-friendly interface. 
  %Uniformly-calibrated photometric zero points are provided for around two-thirds of the data. 
  The background, noise, and sensitivity characteristics of the narrowband $H\alpha$ filter images are outlined.  Typical noise levels in this band correspond to a surface brightness at full $\sim$1 arcsec resolution of around $2\times10^{-16}$ erg cm$^{-2}$ s$^{-1}$ arcsec$^{-2}$.  Illustrative applications of the $H\alpha$ data to planetary nebulae and Herbig-Haro objects are outlined and, as part of a discussion of mosaicking technique, we present a very large background-subtracted narrowband mosaic of the supernova remnant, Simeis 147.  Finally we lay out a method that exploits the database via an automated selection of bright ionized diffuse interstellar emission targets for the coming generation of wide-field massive-multiplex spectrographs. Two examples of the diffuse H$\alpha$ maps output from this selection process are presented and compared with previously published data.  %Down to $i \sim 20$ mag. (Vega system), most stars are also detected in $g$, $r$ and $H\alpha$.  
  %As exposures in the $r$ band were obtained within the IPHAS and UVEX surveys a few years apart, typically, the catalogue includes two distinct $r$ measures, $r_I$ and $r_U$.  
  %The $r$ 10$\sigma$ limiting magnitude is $\sim$21. %Between $\sim$13th and $\sim$19th magnitudes in all bands, the photometry is internally reproducible to within 0.02 magnitudes.  %Stars brighter than $r=19.5$ have been tested for narrow-band $H\alpha$ excess signalling line emission, and for variation exceeding $|r_I-r_U| = 0.2$ mag.  We find and flag 8292 candidate emission line stars and over 53000 variables (both at $>5\sigma$ confidence).  The 174-column catalogue will be available via CDS Strasbourg.
  }


   \maketitle
%
%-------------------------------------------------------------------

\section{Introduction}
\label{sec:intro}

 The stellar and diffuse gaseous content of the Galactic Plane continues to be a vitally important object of study as it offers the best available angular resolution for exploring how galactic disc environments are built, and how they operate and evolve over time.  
 %The optical window in the electromagnetic spectrum remains critical for characterising the properties of stars. 
 For studies of the diffuse interstellar medium (ISM), the optical offers H$\alpha$ -- emission in this strong transition is the pre-eminent tracer of ionized gas.  By definition, the ISM is an extended object that must involve investigation by means of imaging data.  Our purpose here is to present and describe a newly-complete resource that enables this style of astronomy, specifically within the gas- and dust-rich northern Galactic plane.    

%In this era of digital surveys, there is a  growing menu of ground-based wide-field optical broad band surveys covering much of the sky, north and south

There is a history, stretching back over the last century, of comprehensive surveying of the optical night sky.  Until around 1990, much of the wide-area effort depended on photographic emulsions on glass as detectors  \citep[e.g. the Palomar, ESO and UK-Schmidt sky surveys, described by][respectively]{POSSI73, ESOSS74,UKST92}.  The last thirty years has seen a switch to digital detectors that has brought with it the benefits of linearity and increased dynamic range, paving the way for increasingly precise photometric calibration.   Thanks to this change and the advance of data science, the community now has access to a number of wide-area broad band surveys offering images at $\sim$ arcsecond angular resolution and point-source catalogues \citep[e.g. SDSS, Pan-STARRS, DECaPS, Skymapper: see ][]{SDSS, Chambers16, DECaPS,Skymapper}.  

Wide-area narrow band H$\alpha$ imaging, our main focus here, has generally been pursued separately from broad band work \citep[most notably VTSS, SHASSA and WHAM, see ][]{VTSS98,SHASSA01,WHAM03} .  \cite{Finkbeiner03} merged these surveys into a single map covering much of the sky, albeit at an angular resolution limited to 6 arcmin. Inevitably, most H$\alpha$ nebulosity is concentrated in the Galactic plane, along with most of the Galaxy's gas, dust and stars.  Within the plane, the angular resolution needs to be better than this to begin to resolve individual clusters and HII regions that show structure on the sub-arcminute scale.  The UK Schmidt H$\alpha$ Survey \citep[SHS ][]{Parker05}, based on photographic film, has met this challenge in the southern Galactic plane with imaging data of a resolution approaching 1 arcsec.  Before the imaging presented here began to be collected, the same could not be said for the plane in the northern hemisphere.

The focus of this paper is on full coverage of the plane of the northern Milky Way, via digital narrowband H$\alpha$ imaging at $\sim$1 arcsec angular resolution obtained from 2003 up to 2018 using the Wide Field Camera (WFC) on the Isaac Newton Telescope (INT) in La Palma. We showcase the properties of the H$\alpha$ images and point out different modes of exploitation, past, present and future.
Along side this, we also present the IGAPS 5-filter image database, where IGAPS is the acronym for "The INT Galactic Plane Surveys".  IGAPS is the cross-calibrated merger of the two Galactic Plane surveys,  IPHAS \citep[The INT Photometric $H\alpha$ survey of the northern Galactic Plane, ][]{IPHAS}, and UVEX \citep[The UV-Excess survey of the northern Galactic Plane, ][]{UVEX}\footnote{In concept, these two surveys are the older siblings to VPHAS+, the survey covering the southern Galactic Plane and Bulge \citep{VPHAS}.}.  Together these surveys have offered a new mix of narrow-band H$\alpha$ alongside four broad bands spanning the optical.
The IPHAS filters were $r$/$i$/$H\alpha$, %where the narrowband H$\alpha$ filter has a transmission full width at half maximum of 95~\AA\, capturing the neighbouring [N {\sc ii}]~$\lambda\lambda$ 6548, 6584 lines as well as H$\alpha$.  
while the UVEX survey incorporated a repeat in $r$ and $g$/$\URGO$ observations.
The total IGAPS footprint is a 1850 sq.deg. strip along the Galactic Plane defined on the sky in Galactic coordinates by:- $-5^{\circ} < b < +5^{\circ}$ and $30^{\circ} < \ell < 215^{\circ}$.   

\cite{Monguio20} have recently presented the merged and calibrated IGAPS point-source catalogue of aperture photometry derived from the IPHAS and UVEX surveys. The focus of this study was on the extraction of stellar photometric data.  It did not take on characterisation of the extended ionised emission traced by the H$\alpha$ images.  This is the partner paper in which this missing piece is put in place.
%characterising the extended H$\alpha$ emission captured by the IPHAS survey. 

We will outline the world-accessible database of IGAPS images we have set up to hold the H$\alpha$ (and broad-band) data.  It is reached via a website that also provides access to the \citet{Monguio20} point source catalogues. The imagery we have archived incorporates all IPHAS and UVEX observations including a minority that did not meet all the desired survey quality criteria. The majority of the images included benefit from the uniform photometric zero points computed by \citet{Monguio20} in building the IGAPS point-source catalogue.   

The structure of the paper is as follows. In section~\ref{sec:images}, we summarise the relevant features of IGAPS data and our methods and summarise the contents of the image database held within http://www.igapsimages.org/.  This is followed up by an outline of the search tool available for querying the database (section~\ref{sec:website}). Attention then switches to our main focus: the properties and application of the narrowband H$\alpha + [NII]$ imagery.  Section~\ref{sec:ha_properties} sets the ball rolling with an overview of the sensitivity and the nature of the background captured in the narrowband filter.  We then go on, in section~\ref{sec:applications} to highlight two contrasting examples of its exploitation for the science of diffuse nebulae (planetary nebulae and Herbig-Haro objects).  A discussion of techniques for mosaicking the H$\alpha$ data then follows (section~\ref{sec:mosaics}).  Looking to the near future of massively-multiplexed optical spectrographs, we present a method for automated searching of the IGAPS images database for high H$\alpha$ surface brightness non-stellar diffuse-ISM spectroscopic targets 
%and star-free low-flux sky positions
(section~\ref{sec:hagrid}). To round off, we show two contrasting examples of the output from the automated search in section~\ref{sec:hagrid_results}.

\section{Description of the image database}
\label{sec:images}

\begin{table*}\centering
\begin{tabular}{llclcrccc}
\hline
Filter & ING/WFC name & $N$   & exposure & PSF FWHM & sky count & ZP & 5$\sigma$ depth & moon phase \\
          &  &  & (sec)    & (arcsec) & & (mag.)  & (mag.) \\
\hline
$i$  & WFCSloanI  & 83652 & 10 (83\%), and 20 (17\%) & 1.06 &  92 & 26.42 & 20.28 & 0.68 \\
$H\alpha$ & WFCH6568 & 83652 & 120      & 1.20 &  57 & 26.57 & 20.40 & " \\
$r_I$ & WFCSloanR  & 83652 & 30 (85\%), and 10 (15\%) & 1.16 & 164 & 28.19 & 21.37 & " \\
$r_U$ &    "       & 67896 & 30       & 1.18 & 126 & 28.20 & 21.67 & 0.21 \\
$g$ & WFCSloanG    & 67895 & 30       & 1.26 &  61 & 28.71 & 22.38 & "  \\
$\URGO$ & WFCRGOU & 67892 & 120      & 1.48 &  43 & 27.85 & 21.47 & "  \\
\hline
\end{tabular}
\caption{
Properties of the repository contents by filter for the better-quality image sets graded A to C.  The number $N$ specifies the number of CCD images available in the A to C grade range for each filter (D-grade images with identified problems, are not included in the count).  Median values are listed for the full width half maximum of the point spread function (PSF FWHM), the sky background count, zeropoint (ZP), 5$\sigma$ detection limit in the Vega system, and moon phase. The PSF FWHM and the background count are pipeline measures, while the zeropoint and limiting magnitude are based on the uniform calibration \citep{Monguio20}.  Moon phase is given as a fraction, such that 0 corresponds to new moon and 1 to full moon.    The $r_I$ exposure time started out at 10 sec, but was raised to 30 sec from 2004 on, while the $i$ exposure time was increased to 20~sec starting in October 2010.  
}
\label{tab:byfilter}
\end{table*}

A series of papers \citep{IPHAS, IPHASIDR, UVEX, IPHASDR2, Monguio20} has already described the survey data acquisition and pipelining.  So here it is only necessary to repeat pertinent details.

The camera used, the INT's Wide Field Camera (WFC), is a 4-CCD mosaic arranged in an L shape with a pixel size of 0.33\,arcsec/pixel.  Each CCD images a sky region of roughly $11\times22$ sq.arcmin, giving a total combined field per exposure of approximately 0.22 sq.deg.
The five filters used -- $\URGO$, $g$, $r$, $i$, and a narrow-band $H\alpha$ -- have central wavelengths of 364.0, 484.6, 624.0, 774.3, and 656.8 nm respectively. Despite the filter's different naming, the $\URGO$ transmission curve quite closely resembles that of Sloan $u$ \citep{Doi2010}.  We shall refer to the narrow band filter as the $H\alpha$ filter throughout this work, but we note for completeness that the 95~\AA\ bandpass also captures the [NII] 654.8, 658.4 nm forbidden lines typical of HII-region emission.  The numbers of CCD images per filter in the repository are listed along with other performance parameters in Table~\ref{tab:byfilter}.  
%It is also important to be aware that this filter is not a coated-glass interference filter, but is instead based on a cell filled with copper sulphate solution that serves as the red cut-off.  This difference translates into a distinct set of properties compared to that of the longer-wavelength filters.
%The filter set was split between the two surveys such that UVEX covered the blue ($\URGO$, $g$ and $r$), and IPHAS the red ($H\alpha$, $r$ and $i$). 
UVEX and IPHAS $r$ band observations are distinguished by labelling them $r_U$ and $r_I$ respectively.

IPHAS and UVEX shared the same footprint and set of pointings, such that the northern Galactic plane was covered via 7635 WFC pointings, tessellating the survey area with, typically, a small overlap. In addition, each field was observed again, and usually around 5 minutes later, at an offset of +5\,arcmin in RA and +5\,arcmin in Dec in order to fill in the gaps between the CCDs and also to minimize the effects of bad pixels and cosmic rays. As a result, almost all locations in the northern plane have been imaged twice in either survey.   The presence of the $r$ band in both surveys means that there will usually be at least 4 CCD images, uniquely in this band, covering any given position within the footprint.  The only exception to mention is that there is a triangular patch of sky towards $\ell \sim 215^{\circ}$ where there are no UVEX $\URGO$, $g$, $r_U$ images.  

A distinction between the two surveys is that the blue UVEX data were obtained during dark time, while IPHAS observations were generally made with the moon above the horizon.  This difference means that background sky counts are typically higher in the red IPHAS images than they are in UVEX images.  A practical consequence of this is that UVEX $r_U$ data go deeper than IPHAS $r_I$ by 0.3 magnitudes, on average (see Table~\ref{tab:byfilter}).
Another impact is that the background levels in the brighter-time observations in the H$\alpha$ filter can exhibit marked variation, depending on how far away and full the moon is and on cloud cover (see section~\ref{sec:ha_properties}).  These variations, especially when mixed with H$\alpha$ nebulosity, represent a challenge when mosaicking IPHAS data to build large-area images.  How this challenge can and has been met is the subject of section~\ref{sec:mosaics}.

A part of the pipelining procedure for all survey data was to apply a flux calibration based on nightly standard-star observations.  The zero points, in the Vega system, obtained on this basis remain associated with all images in the repository and are stored in every image header as the keyword MAGZPT.
% check early release papers (Eduardo ref ? ask mike where it is defined)
The meaning of MAGZPT is that it is the magnitude in the Vega system of an object giving 1 count per second (it does not fold in the exposure time).
For details of how it is derived see equation 4 in \cite{IPHASIDR}.
Experience showed that broadly consistent and reliable results were obtained for the $\URGO$ and H$\alpha$ filters if their zero points were fixed at a constant offset relative to (respectively) their partner $g$ and $r_I$ frames.  For $\URGO$ this is the only calibration presently available.  %Examples of how this performs were given as Figure 16 in the IGAPS catalogue paper \citep{Monguio20}.
For $g$, $r_U$, $r_I$, $i$ and $H\alpha$, a uniform recalibration of the photometry was undertaken in  preparing the point source catalogue \citep{Monguio20}.
The result of this is that around 2/3 of the images in the repository now carry a revised zero point (PHOTZP header keyword) that rests on a comparison with PanSTARRS $g$, $r$ and $i$ photometry.  This zero point has incorporated the image exposure time, which means it is the magnitude of a source giving 1 count integrated over the exposure. 
Hence, the Vega magnitude of any imaged source can be computed from:
\begin{equation}
    m = -2.5 \log({\rm enclosed\,\,counts}) + PHOTZP 
\end{equation}
where, for a point source, the enclosed counts would be the total counts within a user-specified aperture, after subtraction of an estimate for the enclosed underlying background. 

%For more detail on the broad-band photometric scale, see section 5.3 of the IGAPS catalogue paper \citep{Monguio20}.

The $H\alpha$ filter magnitude scale is not one conventionally defined within the Vega system.  This means we need to define the flux corresponding to the zero of the magnitude scale appropriate to it.  We determine that the zero-magnitude flux entering the top of the Earth's atmosphere within the WFC $H\alpha$ filter transmission profile is 
\begin{equation}
F[m(H\alpha)=0] = 1.57 \times 10^{-7} {\rm ergs\,\,cm}^{-2} {\rm s}^{-1} .
\end{equation}
In the AB magnitude system this is equivalent to a magnitude of 0.328. 

Expressed in point-source magnitudes the bright limit of the images is in the region of 11--12 in $i$ and $H\alpha$, rising to 12--13 in the more sensitive $r$ and $g$ bands.  There were specific observation periods in which WFC electronic issues meant that saturation was reached at appreciably lower count levels than the norm.  For example, there was a particular problem affecting CCD 2 in November 2006 that would push the bright limit fainter by up to a magnitude. Users of the data can track such problems in the images via the count level recorded against the header keyword, SATURATION.

As part of creating the point source catalogue images were graded from best to worst as A++, A+, and A to D. For a definition of the grades see appendix A in \cite{Monguio20}.
A last point to make about the images is to note that, while there is minimal vignetting of CCDs 1, 2 and 4, there is persistent vignetting of the corner of CCD 3 furthest from the optical axis of the WFC (which passes through CCD 4). In Appendix~\ref{app:confidence_map} a typical confidence map is shown to illustrate this. In  Appendix~\ref{app:artefacts}, we briefly outline common artefacts in the image data.

%\begin{itemize}
%    \item Basics of observations (IPHAS and UVEX) -- filters available
%    \item Image dimensions -- served up as individual CCDs --> total numbers in repository per filter.
%    \item Information provided (header content - point to table in appendix?)
%    \item Image quality in U compared to other filters? Instead, just refer to an appendix for itemisation of artefacts, perhaps.  But make mention here of CCD3 vignetting in all cases.
%\end{itemize}

\section{Accessing the images}
\label{sec:website}

All the reduced images from both surveys can be accessed via the website, http://www.igapsimages.org/, hosted by University College London.  Altogether the repository contains 527736 CCD frames, of which 314923 (or 60\%) carry the best-quality A grades and 73097 (14\%) are minimum-quality D graded.  These grade assignments were made at the level of the basic unit of observation in the two constituent surveys -- the consecutive trio of exposures obtained at each sky position.  The details of the grading system at work in evaluating data for the IGAPS catalogue can be found in Appendix A of the catalogue paper \citep{Monguio20}.  A consequence of this approach, also applied to the earlier IPHAS DR2 release \citep{IPHASDR2}, is that the grade assignment, referred to individual CCD frames or even filters, can be pessimistic, as it takes just one substandard exposure in a set of three to pull down the grade for all of them.  In view of this, and the occasional scientific value of maximising the number of images to examine, the decision was taken to retain all D graded data in the repository, alongside the A to C graded exposures.

%{\sl Outline available tool for pulling out data.}  
The images access page within the website offers a search tool that enables users to search for and download images that either overlap a single specified position or occupy a square box of size up to $1\times1$~sq.deg. on the sky. The user can choose the filters of interest and decide whether to omit grade D data.  In response to a query, the tool returns a table listing the images meeting requirement, along with key metadata (grade, seeing, depth, whether calibrated) that can inform the user's final choice of images for download.  For convenience, there is a column of tick boxes in the table, that allows the user to deselect some of the listed images before initiating the download of a gzipped or tarred collection of Rice-compressed (.fz) images.
% refer to thumbnail service as well
%A separate web page is available for extracting thumbnails of the data. The thumbnails can be chosen from three sizes (full CCD, 5 or 1 arcmin). They are generated for a given position in either equatorial or Galactic coordinates. The position can also be specified in the form of a list, with up to 10 shown per output page. 

%In keeping with the way in which the data-taking was split between the IPHAS and UVEX surveys, the $r$-band images from the two surveys are distinguished by naming them $r_I$ and $r_U$ for, respectively, IPHAS and UVEX exposures.  
It is generally the case that a contemporaneous $r_I$ image accompanied an H$\alpha$ and $i$ image of the same pointing.  Similarly the UVEX $r_U$, $g$ and $\URGO$ images were observed as consecutive triplets.  Given that stars are sometimes subject to variability, users of the repository may need to bear this in mind when deciding how to select images for scientific exploitation.

The website also provides a link to a large table of metadata that previews the header information provided with the full set of image profiles.

\begin{figure*}
    \centering
    \includegraphics[width=1.5\columnwidth]{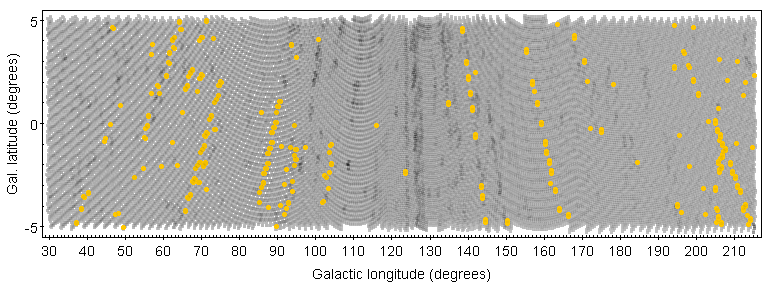}
    \caption{Map of the uniformly-calibrated IGAPS field centres (grey).  A darker grey colour is seen where the database contains more than one uniformly-calibrated exposure.  The orange overplotted points mark the 339 exposure sets (1356 CCD frames), without alternatives in the database, where the $H\alpha$ sky background exceeds 210 counts ($> 95^{{\rm th}}$ percentile). }
    \label{fig:sky_map}
\end{figure*}

%{\sl Copied from above:} The ID for each survey pointing is constructed using four digits, numbering from 0001 and rising with Right Ascension up to 7635, with an "o" straight after in the case of an offset pointing making up the field pair. 

%{\sl Emphasise how $r_u$ pairs with $g$ and $\URGO$ and $r_I$ should be used with H$\alpha$ (and $i$).}

%Will soon ask CDS if they want everything... they already have IPHAS DR2 images.

\section{Properties of the $H\alpha$ images: backgrounds and sensitivity}
\label{sec:ha_properties}

At the outset, the IPHAS survey was allocated time on the expectation the programme could cope with a moonlit sky.  After the first few seasons and some experience had accumulated, the brightest nights were increasingly avoided. Indeed, in the late stages when the acquisition of the blue UVEX filters took priority, dark and grey nights became the norm. The net result is that the median background level among all the uniformly-calibrated $H\alpha$ images is closer to grey, than bright.  Table~\ref{tab:ha_background} provides some numerical detail illustrating the strongly skewed distribution finally achieved.  For comparison with the magnitudes in the table, we mention that the ING exposure time calculator uses 20.6, 19.7 and 18.3 mag arcsec$^{-2}$ to represent dark, grey and bright sky in the $R$ photometric band -- values closely resembling the tabulated 5th, 50th and 95th percentiles.  

Around a third of the images were not passed through uniform calibration: of these (generally inferior) data, just under a third have background levels in $H\alpha$ exceeding the 210 counts marking the 95th percentile of the uniformly-calibrated set of images.  Most of them were obtained early on in the survey, and most have been repeated, resulting in calibrated alternatives.  
Just 339 calibrated $H\alpha$ exposures (1356 CCD frames, or 2 percent of the total) are left without alternatives in the database, where the background count exceeds the 95th percentile of 210 counts.  Where these sit in the survey footprint is shown in Figure~\ref{fig:sky_map}.

\begin{table}\centering
\begin{tabular}{lccc}
\hline
Quantity  & \multicolumn{3}{c}{Percentiles} \\
          &  5  &  50  & 95 \\
\hline  
  & & & \\
Counts   &  27  &   52   & 210  \\
  & & & \\ 
Vega magnitude & & & \\
({\it mag arcsec$^{-2}$}) & 20.6 & 19.9 & 18.3 \\
  & & & \\
Surface brightness & & & \\
($10^{-15}$ {\it erg cm$^{-2}$ s$^{-1}$ arcsec$^{-2}$}) & 0.91 & 1.7 & 7.4 \\
 & & & \\
\hline
\end{tabular}
\caption{
Percentiles of the distribution of narrowband H$\alpha$+[NII] pipeline-computed background levels in uniformly-calibrated images. The sample used here numbers 63956 CCD frames (or 15989 WFC exposures). Not included are the 36808 CCD frames (9202 WFC exposures) for which only the pipeline photometric calibration is available.
}
\label{tab:ha_background}
\end{table}

\begin{figure}
    \centering
    \includegraphics[width=0.85\columnwidth]{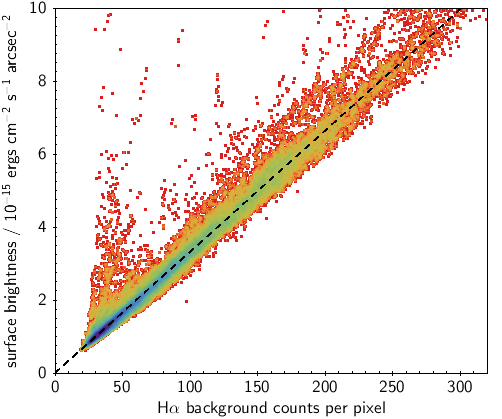}
    \caption{Density plot of background surface brightness in the $H\alpha$ band as a function of measured mean sky level in counts.  The colour scale is logarithmic, with blue representing the highest density of points.  The images used to build this diagram are the IGAPS uniformly-calibrated set, which enables validated conversion of the sky counts to surface brightness via each image's zeropoint.  The black dashed line has a gradient of $3.32\times10^{-17}$ ergs cm$^{-2}$ s$^{-1}$ arcsec$^{-2}$ per count.   }
    \label{fig:sky_counts_SB}
\end{figure}

\begin{figure}
    \centering
    \includegraphics[width=0.9\columnwidth]{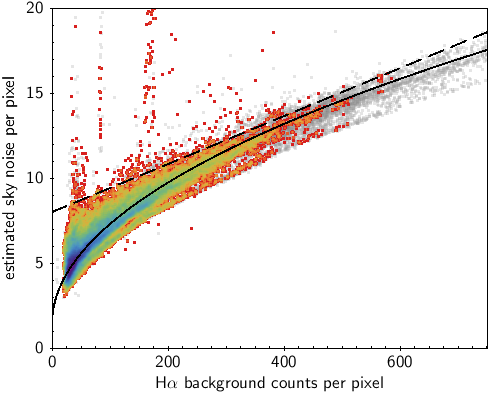}
    \caption{Density plot of the pipeline-measured sky noise in the $H\alpha$ filter as a function of sky level. Both are in units of counts per pixel. The coloured data shown, and plotted on a logarithmic density scale, are drawn from the IGAPS uniformly-calibrated set.  The solid black line is the fit discussed in the text, given as equation~\ref{eq:noisefit}, that shows the sky noise increases mainly as the square root of the sky level.  The lighter grey data (underneath) are from images with zeropoints that were not passed through uniform photometric calibration.
%    The straight dashed line is a more conservative upper limit.   
    }
    \label{fig:sky_noise}
\end{figure}

%By the time the survey completed, a dark sky background applied in around 5 percent of all calibrated frames. Figure~\ref{fig:sky_counts_SB} shows how the measured background count maps onto surface brightness, as computed from the available zeropoints.  The minimum sky brightness at H$\alpha$ for new moon was $\sim$20 counts. In around 5 percent of exposures it was very high, at above $7.5\times10^{-15}$ ergs cm$^2$ s${-1}$ arcsec$^{-2}$.  This level maps onto 18.3 mag arcsec$^{-2}$, identified as 'bright' by the Isaac Newton Group exposure time calculator. 

Figure~\ref{fig:sky_counts_SB} shows how the recorded sky background count levels convert into narrow-band $H\alpha$ surface brightness, via the set of zeropoints, PHOTZP. The plot includes only data that have passed through the uniform photometric calibration.  Most of the data conform reasonably well to a linear trend such that 1 count per pixel corresponds to $3.32\times10^{-17}$ ergs cm$^{-2}$ s$^{-1}$ arcsec$^{-2}$ (or 5.9 Rayleighs)\footnote{At H$\alpha$, 1 Rayleigh is equivalent to $5.67\times10^{-18}$ ergs cm$^{-2}$ s$^{-1}$ arcsec$^{-2}$}.   The absolute minimum sky brightness measured in any calibrated frame is $6.6 \times 10^{-16}$ ergs cm$^{-2}$ s$^{-1}$ arcsec$^{-2}$ ($\sim 120$ Rayleighs), but it is commonly more than twice this.  Sky transparency necessarily influences the behaviour.  At times of reduced transparency the sensitivity suffers, driving data points from affected nights onto steeper linear trends in Figure~\ref{fig:sky_counts_SB}.

Figure~\ref{fig:sky_noise} is a plot of the pipeline measurement of sky noise versus the estimate of sky background, both in counts per pixel. For around a half of the uniformly calibrated frames, the noise is limited to under 6 counts per pixel (or $\sim 2 \times 10^{-16}$ ergs cm$^{-2}$ s$^{-1}$ arcsec$^{-2}$), and it is extremely rare that the sky noise is more than $\sim$12 counts per pixel. 
A simple expectation for the form of the relation between sky noise and level is
\begin{equation}
    N_{bg} = ( N_{RN} + \sqrt{C_{bg}} ) / \sqrt{n_{pe}}
    \label{eq:noise}
\end{equation}
where $N_{bg}$ and $C_{bg}$ are respectively the sky noise and sky level, $N_{RN}$ is the CCD read noise, also in counts per pixel, and $n_{pe}$ represents the effective number of pixels over which the sky statistics are measured. Basically, we expect the noise to be the sum of a constant and a Poisson component.
%averaged over a certain pixel area. 
In practice, the sky level is determined numerically by the pipeline using two-dimensional non-linear background tracking across the full CCD at a superpixel level. After the background fit has been subtracted from an image, the sky noise is calculated iteratively from the clipped median absolute deviation (MAD) of the residuals.
%(which are assumed to be normally distributed). 
%\textbf{JED Q: why scaled?  ...sure about this? Robert A: thats what Mike wrote in his email. I think he refers to the factor 1.48, which converts the MAD to $\sigma$ for a normal distribution.}
%differential background subtracted image.

As the readnoise for the WFC is known to be $N_{RN}=2.37$ counts, we only need to fit one parameter of this function. Before the fitting, outliers above the dashed line in Figure~\ref{fig:sky_noise} were removed. The fit was performed using an iterative 3-$\sigma$ clipped Levenberg-Marquardt algorithm using the median and MAD in place of mean and $\sigma$. The result of the fit is
\begin{equation}
    N_{bg} = ( 2.37 + \sqrt{C_{bg}} ) / \sqrt{2.88} = 1.4 + 0.589 \sqrt{C_{bg}}
    \label{eq:noisefit}
\end{equation}
Allowing an additional fit parameter, namely a multiplicative factor for the $\sqrt{C_{bg}}$ term, does not lead to a statistically significant improvement in the fit, and the factor is found to be very close to 1. Hence we are confident that equation \ref{eq:noise} accurately describes the distribution.

In presenting the Schmidt H$\alpha$ Survey, \cite{Parker05}, compared its sensitivity with IPHAS and other available H$\alpha$ wide-area surveys.  In their Table 1 it was estimated that the depth reached by the narrowband (IPHAS) data presented here is $\sim$3 Rayleighs.  This is a surface brightness of $1.7 \times 10^{-17}$ ergs cm$^{-2}$ s$^{-1}$ arcsec$^{-2}$ or, as required by the mean calibration illustrated in Figure~\ref{fig:sky_counts_SB}, the equivalent of around a half count per pixel in a 120~sec exposure at the typical 1 to 1.5 arcsec seeing. The on-sky solid angle that the \cite{Parker05} estimate refers to was not made explicit -- clearly it cannot be 1 arcsec$^2$. %The estimate made by \cite{Parker05} clearly demands more detail.
%In particular it is necessary to specify the level of significance (in sigma) and the sky area over which the limit has been calculated. 
If we assume $C_{bg} = 0.5$ counts pixel$^{-1}$ at 3 $\sigma$, $N_{bg} = 0.5 / 3$, we can use equation \ref{eq:noise} to calculate the pixel area needed to achieve a sensitivity of 3 Rayleighs. We obtain $18.5 \times 18.5$ pixels or $6.1 \times 6.1$ arcsec.

Used at full seeing-limited resolution, the $H\alpha$ data provide safe detection of raised surface brightness due to diffuse-ISM emission at the level of a few times $10^{-16}$ ergs cm$^{-2}$ s$^{-1}$ arcsec$^{-2}$, or 50--100 Rayleighs. That this is so will be demonstrated by different means in section~\ref{sec:hagrid}.

%rg-end

\section{Exploitation of the $H\alpha$ images}
\label{sec:applications}

%Preamble, briefly outlining the past in order to point to obvious existing publications.
The search for and characterization of extended nebulae was an original goal of the IPHAS component of the merged IGAPS survey. Initially this mostly concentrated on planetary nebulae (PNe).
%Examples of discoveries are the rare quadrupolar PN \citep{Mampaso2006}, the Necklace nebula \citep{Corradi2011}, and a PN around the classical nova, V458 Vul \citep{2008ApJ...688L..21W}. The images have also enabled the tracing of fainter interactions between PNe and the ISM \citep{Wareing2006}.
Discoveries of examples of those alternative products of end-state stellar evolution -- supernova remnants -- have also been found and recorded \citep{Sabin2013}.  

Below we present brief discussions of how nebulae can be found and studied, taking the  contrasting examples of new PNe and -- from the more obscured first phase of the stellar life cycle -- a Herbig Haro object. 

%Current sub-headings are entirely open to negotiation. % Section to include some nice image cut-outs to illustrate discussion.

\subsection{Planetary nebulae}
\label{sec:PN}

 %{\bf A catalogue of extended PNe found from a visual search of binned mosaics was presented by \cite{Sabin2014}. Rhys et al. with lots of help from Laurence.}

Planetary nebulae (PNe) are the end-products of low and intermediate mass stars ($\sim$1--8M$_{\odot}$) where a hot central star fully ionises its surrounding shell leading to a glowing nebula. These objects are ideal tools to study the later stages of stellar evolution as most stars in our Galaxy will go through this phase. We can access information related to their physical characteristics via plasma diagnostics (density, temperature, velocity and so on) and the chemical composition allows the measurement of their impact on the chemical enrichment of any given galaxy. 
%The more that is known about PN then the better is our understanding of stellar evolution. 
However most of the known PNe are 
%generally 
bright and/or nearby as the faint ones have been ignored or are out of reach due to observational constraints. An aim of IPHAS was to perform a near complete census of the PNe in the northern Galactic Plane, where a higher concentration of ionised sources co-exists with high extinction. Progress towards this goal has been described by \cite{Sabin2008} and \cite{Sabin2014}.
%Thanks to the detectors sensitivity and resolution (0.33 arcsec pixel$^{-1}$ plate scale), combined with the detection methods used, it was possible to scan the northern Galactic Plane including the relatively unexplored Anticentre region.

\subsubsection{Detection Methods}
Depending on the size of the sources two methods were adopted. On the one hand, compact or point-source PNe have been selected based on H$\alpha$ excess as measured and recorded in photometric catalogues, and cross-checked against IR photometry \citep[e.g. 2MASS, see][]{Viironen2009}.  On the other hand, for the case of extended PNe, a mosaicking process was developed using 2$^{\circ}$\,$\times$\,2$^{\circ}$ $H\alpha-r$ mosaics with 5$\times$5 pixels and 15$\times$15 pixels binning factors \citep[corresponding to $\sim$1.7 and $\sim$5 arcsec effective pixel sizes respectively, see][]{Sabin2008}.  
The coarser binning was mainly used to detect large and low-surface brightness nebulae, while the lower binning was aimed at detecting smaller nebulae hidden in crowded stellar fields. In the latter case we also used the technique adopted in the southern MASH survey \citep{Parker2006,Miszalski2008} where RGB composite imagery (H$\alpha$, {\it r'} and {\it i'} filters) is used to distinguish stars from diffuse ionised nebulae.\footnote{With advances in the use of machine learning, the classifications of objects in the HASH database can now be automated \citep{Iskandar2020}} Always, new objects were 
confirmed by independent eyeballing of the data by several team members. 

This task was found to be ideal for undergraduate projects and two example discoveries are shown below.

\subsubsection{New Discoveries}

Hundreds of new PN candidates were found as a result of the visual search with the mosaics and a first large spectroscopic follow-up involving no less than nine world wide telescopes/instruments ranging from 1.5m (ALBIREO spectrograph at the Observatory of Sierra Nevada, Spain) to 10.4 m (GTC-OSIRIS spectrograph at the Observatorio del Roque de los Muchachos, Spain) was conducted \citep{Sabin2014}. We identified 159 objects as True (113), Likely (26) and Possible (20) PNe and unveiled a large range of shapes (mostly elliptical, bipolar and round PNe) and sizes (up to $\sim$8 arcminutes). 
%We also note that 83 candidate PNe were found by the means of the photometric search.

Among all the newly detected PNe, some caught our attention first due to their outstanding morphology and then due to their interesting characteristics revealed by subsequent deep analyses. Notable examples include: the knotty bipolar IPHASX J194359.5+170901, now known as the Necklace Nebula \citep{Corradi2011}; the quadrupolar IPHASX J012507.9+635652 \citep[alternately named {\it Pr{\'i}ncipes de Asturias by}][]{Mampaso2006}; IPHASX J052531.19+281945.1 at large galactocentric distance \citep{Viironen2011}; and finally, the PN discovered around the nova V458 Vul \citep{2008ApJ...688L..21W}.  IPHAS, and now IGAPS, also allows us to target the particular group of PNe interacting with the interstellar medium (ISM). Those objects which are slowly diluting in the ISM are particularly difficult to detect most of all in their late phase. In this case IPHAS imaging data can reveal such faint material \citep{Sabin2010}.

\begin{figure}[h]
    \centering
    \includegraphics[width=0.49\textwidth]{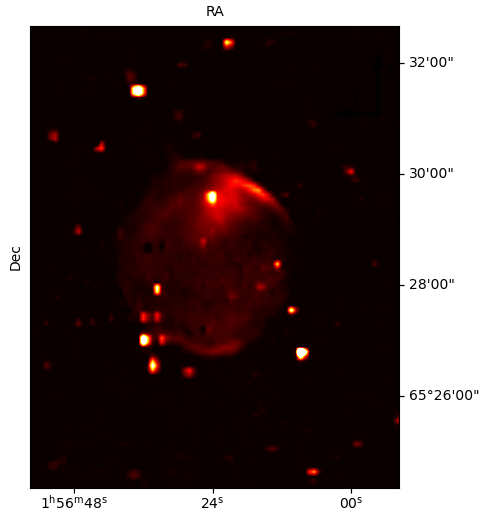}
    \caption{\object{IPHASXJ015624.9+652830}: this is a cutout from one of the 2 by 2 degree binned H$\alpha$-r mosaics used for searching for new PN. }
    \label{fig:bubble}
\end{figure}

\begin{figure}[h]
    \centering
    \includegraphics[width=0.49\textwidth]{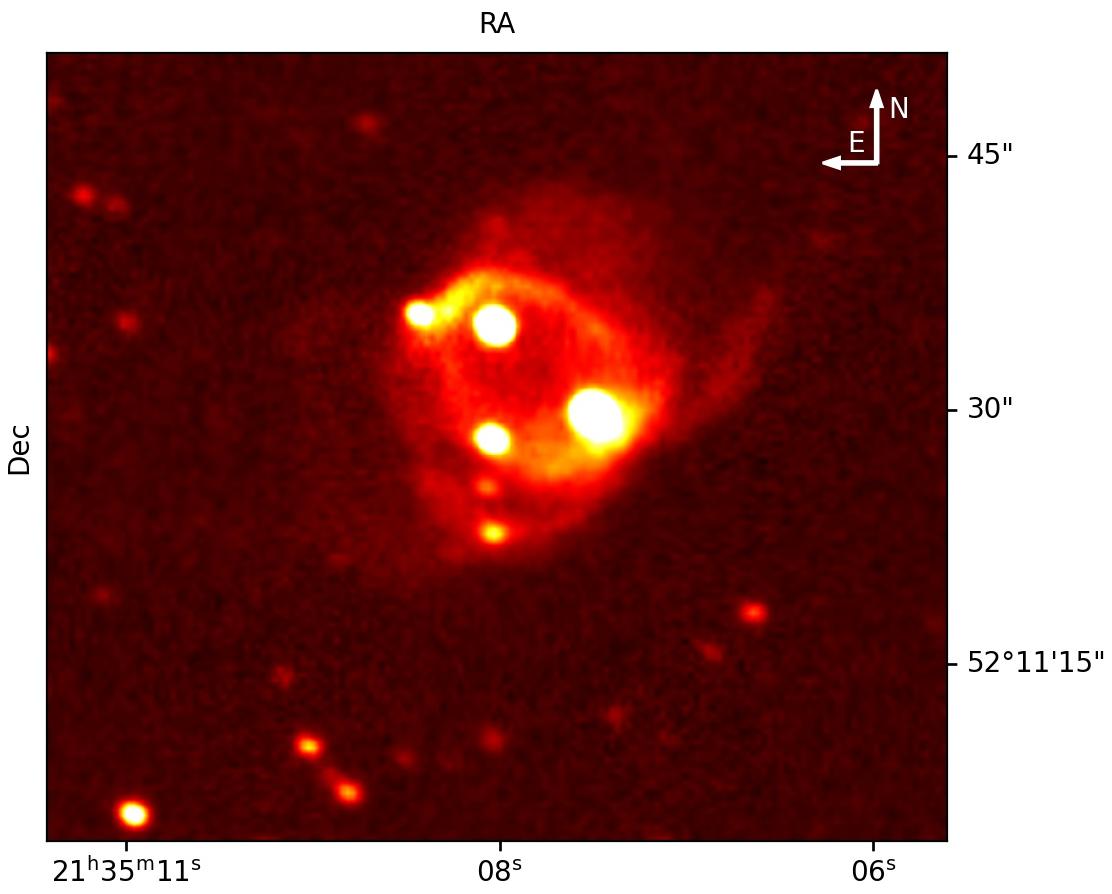}
    \caption{\object{PNG 95.3+0.2}: this is a cutout from one of two CCD frames downloaded from the IGAPS image archive.  The alternative available image of the PN (not shown) happens to sit adjacent to a 'bad line' in the CCD.  It is precisely to mitigate against this kind of problem that the survey strategy was to collect data from paired offset pointings. The object IPHAS J213508.15+521128.0, discussed in the text, is the lower left of the three bright stars in the central region of the nebula.}
    \label{fig:antonio}
\end{figure}

We show in Figures~\ref{fig:bubble} and \ref{fig:antonio}, two of the many new PNe that have been found and have not yet appeared in the refereed literature.
%by final year undergraduate project students, 
Most of the new finds are published in \cite{Sabin2014}. The first object  IPHASXJ015624.9+652830 (PNG 129.6+03.4), shown in Figure~\ref{fig:bubble}, came from parallel searches of the $15\times15$ pixel binned $H\alpha - r$ (5 arcsec per pixel) mosaics, carried out by team members and undergraduate students supervised by them in 2007. This object can be seen in the H$\alpha$ images but is easily missed, yet it stands out very clearly in the binned difference images as the only object in an otherwise blank frame. This demonstrates the usefulness of this method for the discovery of new objects, especially those that, like this one, are only detectable in optical wavebands. 

%The PN in Figure \ref{fig:bubble} was found by project students Matina Mitchell and Paul May and their supervisor Rhys Morris at the University of Bristol and by another author of this paper, Laurence Sabin, around the same time in 2007. The image is a cutout from the 5 x 5 pixel binned H$\alpha$-r mosaics.<p>

%Figure \ref{fig:antonio} is PNG 95.3+0.2 found by Alba Fernandez in 2007 working at the IAC. This image is a cutout of an image downloaded from the IGAPS website. As part of the survey design this object was imaged twice, in a standard and an offset pointing, which works well as one of the images has a defect running through this object. 

% Text from Antonio - start

The second nebula, PNG 95.3+0.2, shown in Figure~\ref{fig:antonio}, was discovered using the same search method by \cite{Fernandez2007}.   Subsequent high resolution (0.6 arcsec seeing) imaging with the NOT telescope on the night of 4th September 2007, through H$\alpha$, [OIII] 5007{\AA}, and [NII] 6583{\AA} filters further clarified an intricate morphology with a bright, roughly elliptical central shell, and a pair of fainter twisted protrusions that show up especially well in the [NII]-only image. Since the IPHAS H$\alpha$ filter bandwidth also incorporates this line, Figure~\ref{fig:antonio} is a composite image, showing both the [NII]-dominated outer filamentary structure and the H$\alpha$ dominated inner ellipse.

PNG 95.3+0.2 is now listed in the HASH Catalogue of \cite{Parker2016}.  It is associated with an infrared (WISE, and 90$\mu$m AKARI), and 1420 MHz radio source \citep{Taylor2017}. According to \cite{Anders2019}, the $G \simeq 17.1$ mag star near the geometrical centre, IPHAS J213508.15+521128.0, or {\it Gaia} EDR3 2171830374492778880, is a distant, reddened, and apparently relatively cold star (D = 5.5$\pm$ 0.9 kpc, $A_V = 3.9 \pm 0.2$~mag, $\mathrm{T_{eff}} = 4800 \pm 260$~K).
However, our evaluation of the IGAPS broad-band photometry \citep{Monguio20} is that the available magnitudes are also consistent with this object being a much hotter, even more extinguished star.  The other two, brighter stars embedded in the nebula are located in the foreground at much more secure parallax-based distances of 0.9 and 1.6 kpc: in the \cite{Anders2019} database, they too are assigned low $\mathrm{T_{eff}}$ values, incompatible with those of a hot PN central star. \cite{Vioque+2020} combine IPHAS, 2MASS and WISE data in a search for new Herbig Ae/Be stars and list IPHAS J213508.15+521128.0 as a non-Herbig AeBe, non-pre Main Sequence, and non-classical Be star -- nor do they confirm an association with a PN (their FPN flag is empty). The WISE source, detected in the four bands and centered at 1.1 arcsec from IPHAS J213508.15+521128.0, shows red IR colours like known PNe, while the spectral energy distribution of the star, built from Pan-STARRS, 2MASS, ALLWISE, and AKARI data, is typical of a reddened star up to the WISE W3 (12$\mu$m) band. Beyond that, a strong IR excess appears up to 90$\mu$m, and points to a physical association of IPHAS J213508.15+521128.0 with the nebula. If that is the case, the apparent nebular size of around 30 arcsec would imply a rather large, evolved nebula, 0.8 pc in diameter, and also suggest the existence of a hidden hot star (a binary?) in the surroundings. PNG 95.3+0.2 is an appealing example of an IPHAS extended object with plenty of online, publicly available information, that nevertheless deserves further dedicated observations including careful quantitative spectroscopy to pin down the central star.
% Text from Antonio - end

Basic confirmation spectra exist for both the above nebulae. The objects have also been independently discovered more recently by amateur astronomers. The first object,  IPHASXJ015624.9+652830, has also come to be known as Ferrero 6, Fe6, PN G129.6+03.4, while the second, PNG 95.3+0.2, is also known as StDr Objet 1. Full details on both can be found in the HASH database \citep{Parker2016}.

\subsubsection{Previously known PN}

The survey is also useful for the re-analysis of already known PNe. IPHAS images can unveil new faint structures associated with known PNe that have hitherto evaded detection.  A clear example of this is the detection of the extended tail of the known Sh 2-188 by \citet{Wareing2006} which enabled the reevaluation of its full extent.  A different application has recently been presented by \cite{Dharmawardena21}: with a view to appraising different methods of determining PN distances, they have collected IPHAS H$\alpha$+[N~{\sc ii}] aperture photometry fluxes for 151 previously known nebulae as well as for 46 confirmed or possible PNe that had been discovered by the IPHAS survey.
\\

%\subsection{Supernova remnants}
%\label{sec:SNR}

%If we don't want to have this section, some comment on SNRs (including reference to the Sabin et al 2014 paper) could be folded into PN section.

%\subsection{Old nova shells}
%\label{sec:novae}

%Simo... ?

%\subsection{HH objects}
%\label{sec:HH}

%For Jochen and Bringfried to design in first instance.  Do we want to broaden this to 'star formation' more broadly to take in HII regions (mix of new and old, inc reference back to the Cygnus 'proplyds' paper)?

\subsection{HH-objects}
% (Bringfried Stecklum and Jochen Eislöffel)\\
During their growth, young stellar objects (YSOs) eject a fraction of the in-falling matter at high speed via bipolar jets and outflows. Their shock fronts, delineated by line emission, particularly in H$\alpha$, [SII], and [OI], are called  Herbig-Haro objects (HHOs, \citealp{1950ApJ...111...11H,1952ApJ...115..572H}). HHOs not only trace the presence of young stars, but can also serve as a record of their accretion history. The kinematics of HHOs, derived from proper motions (PMs) and radial velocities (RVs) allow a kinematic dating of the ejection event. Moreover, such data provide information on the inclination $i$ of the circumstellar accretion disk. Constraining the latter is crucial for the analysis of YSO spectral energy distributions using radiative transfer modelling.
\begin{figure}
%\begin{figure}[h]
    \centering
    \includegraphics[width=0.9\columnwidth]{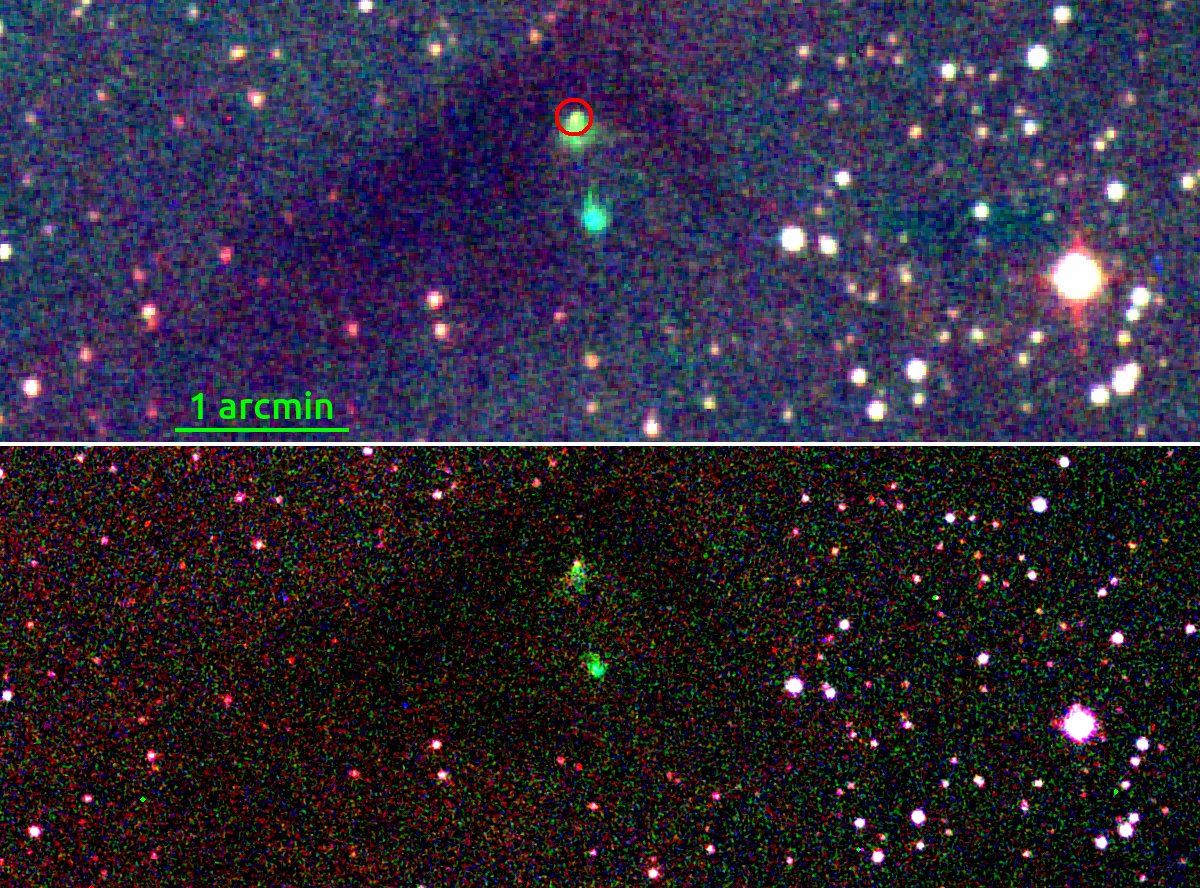}
    \caption{\textbf{Top:} RGB image (epoch 2007) of the dark cloud Dobashi 3782, based on $I, H\alpha$, and $[SII]$ frames taken with the Tautenburg Schmidt telescope. It hosts the YSO \object{IRAS 01166+6635}, marked by the circle, which drives a jet that excites a compact HHO south of it (centre). It appears greenish-blue in this RGB representation because of its strong H$\alpha$ emission. \textbf{Bottom:} IGAPS RGB image (epoch 2013), based on $i$, $H\alpha$, and $r_I$ frames.  The YSO, as well as the HHO, appears green because it is not clearly detected in either $r_I$ or $i$.  The HHO is just resolved at no more than 2 arcsec across. 
    %\textbf{JED -- how about we add an IGAPS $H\alpha$ cut out here, from a contrasting epoch?}
    }
    \label{fig:hho}
\end{figure}

The potential of IGAPS for such studies is illustrated by the example of a hitherto unknown HHO, driven by IRAS 01166+6635. This low-mass YSO \citep{2008AJ....135.2496C} is emerging from the small dark cloud, Dobashi 3782, situated at a kinematic distance of 240\,pc \citep{1989A&AS...80..149W}. Narrow-band imaging performed in 2007 with the Tautenburg Schmidt telescope revealed a compact HHO south of the YSO within $\sim$1 arcsec of the position 
% from 2003 IPHAS
RA 01:20:02.9, DEC +66:51:00 (J2000) (Fig.\,\ref{fig:hho}).  The estimated extinction out to the distance of this cloud is about $A_V \sim 2.5$, averaged across a few arcminutes \citep{Sale14}.  The extinction towards the optically-faint YSO is without doubt much more. 
%The HHO possibly escaped previous detection due to its stellar appearance. 
At the position of this YSO there are eight entries in the IGAPS catalogue \citep{Monguio20} within a radius of 5 arcsec. Six of the eight are \Ha only sources. 

Blinking with the POSS1 red image showed evidence for proper motion within $\sim$50 years. Thus, in order to establish its kinematics, $H\alpha$ and $R$-band frames have been secured for as many epochs as possible. Four $H\alpha$ frames with grade A quality (r367494 and r367497 obtained 2003, r1018994 and r1018997 obtained 2013) were retrieved from the IGAPS image server. These are supplemented by archival Tautenburg $H\alpha$ images (2007, 2012) and two frames taken in 2020 with the new TAUKAM instrument \citep{2016SPIE.9908E..4US}. The POSS 1 and 2 $R$-band images (1954, 1991) were added as well, using both plate digitisations from STSci and SuperCosmos. Before deriving the HHO positions by fitting its image profile, all frames were tied to the {\it Gaia} EDR3 astrometric reference system \citep{2020yCat.1350....0G}. 
\begin{figure}[h]
    \centering
    \includegraphics[width=0.9\columnwidth]{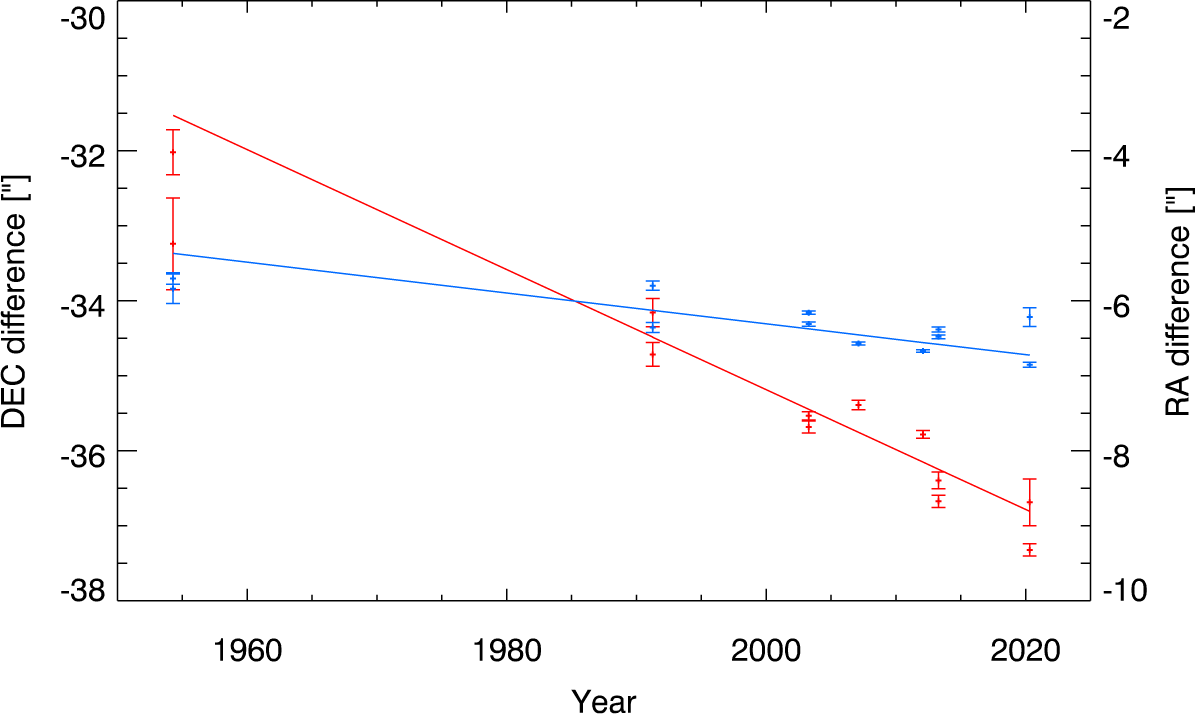}
    \caption{HHO position displacement with regard to the ALLWISE position \citep{2014yCat.2328....0C} over time (red - DEC, blue - RA). The respective regression lines are shown as well.}
    \label{fig:hho_pm}
\end{figure}

The coordinate offsets of the HHO relative to the driving source for the various epochs are shown (Fig.\,\ref{fig:hho_pm}), along with their respective linear fits. For the distance given above, the DEC slope (red) corresponds to a velocity of $-91.0\pm3.5$\,km\,${\rm s^{-1}}$. Assuming constant speed, this implies a kinematic age of $\sim$430\,yr. It is likely a major accretion event happened around that time which induced jet strengthening. Accounting for the position angle of the HHO movement of $190^{\circ}$ (measured from N through E), a total PM speed of $92.4\pm3.5$\,km\,${\rm s^{-1}}$ can be derived. With the help of an RV estimate of $-74.8\pm4.6$\,km\,${\rm s^{-1}}$, obtained by low-resolution spectroscopy of the HHO using the Nasmyth spectrograph at the Tautenburg telescope, the inclination of the velocity vector follows as $51^{\circ} \pm 2^{\circ}$. This intermediate inclination is consistent with the cometary appearance of the YSO in the optical.

This example shows that the IGAPS $H\alpha$ line emission images provide an excellent means for the detection of HHOs.  In this instance, neither the driving YSO nor the HHO are detected even in $i$-band, and yet the detection of the HHO in $H\alpha$ is clear, at around 30 counts above background (depending on the seeing).   Moreover, the
$1$-arcsec resolution of IGAPS images allows for good position measurements of these small (but extended) nebulous objects. Together with the availability of repeated IGAPS observations, with several years of epoch difference, precise proper motion measurements are possible, from which information about the physics of the HH flows, as well as their driving sources, can be obtained.

%\subsection{The diffuse interstellar medium}

%{\sl This section may disappear, or get rolled into previous section - depending on how things balance for content.}
%H-alpha imagery traces ionised ISM.  Talk about striations in/about Cygnus-X region and possible link up to Federrath/Aris explanation of such features in terms of interstellar B field.  Show examples.

\section{Image mosaics}
\label{sec:mosaics}

% The rebuild of the S147 large area mosaic as a background-cleaned H-alpha mosaic by Charlie Roe could be a central item here, with the mosaic itself provided as supplementary material.  One or two details could appear as a figure.  This could be a place to discuss the generalities of doing this (Nick Wright). 

% First draft complete (NJW, 26 Jan), need to do some editing and tightening and get the references in properly, but I think the overall content is broadly right.

\begin{figure*}
    \centering
    \includegraphics[width=1.6\columnwidth]{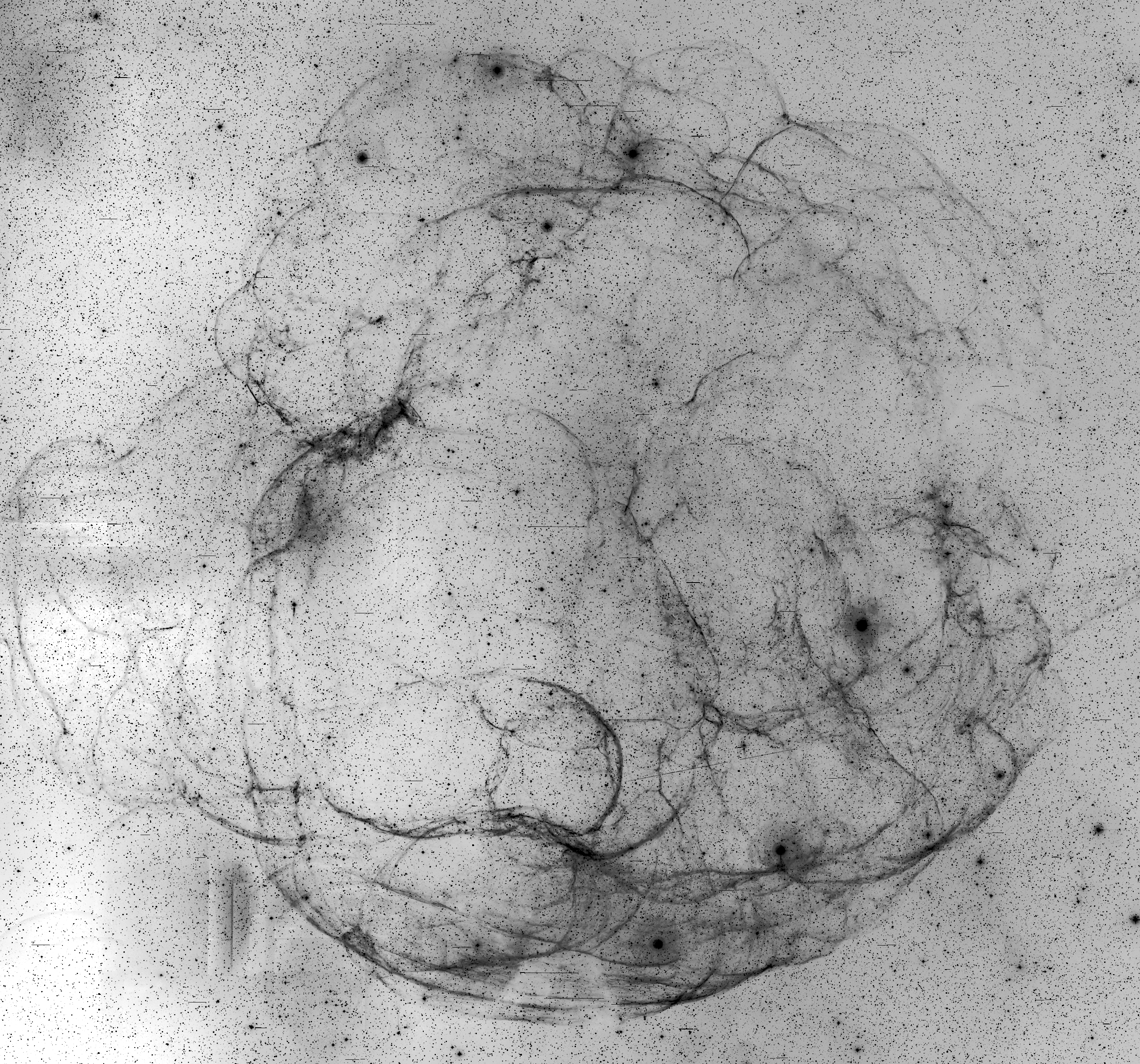} % orig 7.2M
    \caption{Supernova remnant, S147. This is a full resolution background-corrected mosaic made from the $H\alpha$-filter data (no $r$ subtraction). Approximate image dimensions are $4.2 \times 3.6$ sq.deg.  North is up and east is to the left.  The greyscale is negative such that the brightest emission is darkest.  Compared to the earlier mosaic appearing as Figure 19 in \cite{IPHAS}, based on $H\alpha - r$ data, with background fitting and removal, there are fewer artefacts thanks in large part to the incorporation of better re-observations.  
    %There are some improvements in the evenness of the background level also. 
    Much of the faint structure left, such as the ragged diffuse emission in and below the centre of the remnant,is real.  
%    The right-hand panel superposes in red the down-sized {\sc HaGrid} selection of sky positions for follow-up spectroscopy (described in Section~\ref{sec:hagrid}) on the \cite{Finkbeiner03} low resolution H$\alpha$ map, shown in linear grey scale.  In both panels, Galactic latitude increases up, while Galactic longitude increases to the left.
    }
    \label{fig:s147_mosaic}
\end{figure*}

The background of astronomical images 
%taken with ground-based telescopes 
would ideally be flat and dark. In reality the background in images from ground-based telescopes varies due to the interplay of different sources (e.g., airglow, moonlight) contributing varying levels of unwanted light.

To begin to tackle this securely, when working with the standard image data reduction available in the database, 
%In addition to the standard image data reduction process, 
we recommend using the available confidence maps to reject pixels with confidence levels less than 90-95\% typically.  But we note that the level can vary across the sky and between filters, usually requiring a lower confidence level threshold for the $g$ filter and especially $\URGO$.

The {\sc Montage} software\footnote{Available from http://montage.ipac.caltech.edu/ } can be used to effectively mosaic images in a given filter, re-projecting images as needed to a single-projection algorithm and direction, adjusting the background levels in overlapping pairs of images to produce a smooth mosaic over a large area.

In addition to uniform background light sources, the $i$-band can sometimes be beset by fringing that originates from the airglow OH lines interfering via internal reflections in the CCD chips. This wave-like structure contributes $\sim$2\% of the total counts in some images \citep{Irwin01}. It is almost entirely removed in the CASU pipeline (who use a library of $i$-band fringe frames from other INT WFC observing runs), although some will remain at the $\sim$0.2\% level due to night-to-night variations. When $i$-band data is mosaicked, overlapping fringing is occasionally exaggerated and can remain visible in some mosaics.

As noted in Section~\ref{sec:ha_properties}, IPHAS observations were at first carried out at any level of moon brightness throughout the Galactic plane season. Observations during bright time were soon found to exhibit varying levels of background counts in the form of a small but noticeable gradient across each CCD (leading in later seasons to tighter moon phase and distance requirements). Ultimately, $\sim$8\% of all IPHAS images were taken under such conditions.

Moonlight affects IPHAS images through both scattered light across the night sky and a component that reflects off the inside of the telescope dome and across the CCD array. The resulting illumination is therefore not necessarily uniform across all four CCDs, and requires a CCD-by-CCD solution. Its character is also influenced by the phase of the moon, its altitude above the horizon, angular separation from the pointing of the telescope, and the extent (and position) of cloud cover across the sky. These relatively small gradients can be exacerbated by mosaicking CCD images over large areas of the sky (many degrees), becoming a significant issue in the production of large mosaics.

% need a number here to indicate the range of moonlight gradients across a CCD. Charlie's thesis suggests "a couple of counts to an image, up to the order of ~2000 counts", but that doesn't give us a range of gradients.

The recommended solution to removing the moonlight and achieving a flat and dark background is to model and fit the background gradient for each CCD. Since the $r$ and $H\alpha$ band images contain nebulosity that could affect the fit, we recommend fitting the background gradient to $r-H\alpha$ images (after scaling the images to correct for their different exposure times), since both filters contain the H$\alpha$ and forbidden [N{\sc ii}] lines that typically dominate diffuse astronomical emission. Binning the image into $100 \times 100$ pixel bins and taking the median pixel value in each bin provides a simple method to measure the background level in that bin. A two-dimensional gradient of the form $Z = Ax + By + C$ can then be fit to the data, where $A$, $B$ and $C$ are free parameters, using, for example a Markov Chain Monte Carlo simulation and the Python code emcee \citep{emcee13}. More complex models have been tested (including Fourier transform techniques), but none were found to provide a significant improvement. In short, the two-dimensional gradient method will prove effective in the majority of cases.

Some care should also be applied when very bright stars fall on (or near) one of the CCDs, as saturation, atmospheric and lens effects can heavily affect an image and the model fit to it. Identifying and excluding a magnitude-dependent radius around such stars using the Tycho-2 catalogue of bright stars \citep{TYCHO2} proved effective for overcoming these problems.

As an example of the potential of IGAPS image mosaicking, we present a large, $4.2 \times 3.6$ degree $H\alpha$ mosaic of the supernova remnant Simeis 147, produced using the techniques described above.
Simeis 147 (hereafter S147) is otherwise known as \object{SNR G180.0-01.7}, Shajn 147, or Sharpless 2-240. It is a large, faint, late-stage remnant located just below the Galactic Anticentre. It was discovered in 1952 and lies at a distance of 0.8--1.6 kpc \citep{Gvaramadze06}, on the near side of the Perseus spiral arm. The SNR consists of numerous filaments embedded in large-scale diffuse emission. The east and west edges of S147 show signs of blow-outs, the southern edge shows a sharp boundary, with a less regular one in the north. The undistorted appearance of the SNR may partly be due to it expanding into a region of space already partly cleared by a previous supernova \citep{Gvaramadze06}. Regardless, examples of large and pristine SNRs are rare and observations of them can be important for constraining hydrodynamic simulations of their expansion and structure.

Figure~\ref{fig:s147_mosaic} shows the full mosaic constructed as outlined above.  The challenge of this object is its great size, allied with very intricate and sometimes very faint small scale detail.  The number of individual CCD frames included is in the region of 250.  The full-resolution S147 mosaic (199 MB) itself is provided as a fits-formatted file attached to this paper as supplementary material.  
%\textbf{JED: we need to decide how to balance figure content on S147 between this section and section 7.3}

% Figure 51 from Charlie's thesis? And perhaps an inset showing the zoom-ins from Figure 53?

\section{Nebular target selection for massive-multiplex spectroscopy}
\label{sec:hagrid}

%While the last decades saw the emergence of wide-field photometric digital sky surveys on 2m class telescopes, 
The next decade will see an increase in large, multi-object digital spectroscopic surveys on 4m class telescopes. In drawing up target lists, these surveys will make use of the data that has been acquired by wide-area digital photometric surveys, like IGAPS. Two examples due to start soon are the WEAVE survey on the 4-metre William Herschel Telescope (WHT) of the Roque de los Muchachos Observatory in La Palma \citep{Dalton2020} and the 4MOST survey on the 4-metre VISTA telescope operated by ESO at Paranal \citep{Dejong2019}.  Both facilities will collect on the order of 1000 targets per pointing, and a major science driver for both is Milky Way science in the {\it Gaia} era.

Massive-multiplex spectroscopy requires informed target selection.  The IPHAS $H\alpha$ images, in particular, can characterise the diffuse sky for studies of the ionised ISM. 
Here, we outline a software method -- named \hagrid -- aimed at doing this through the interrogation of H$\alpha$ images.  The positions generated by \hagrid will be used in constructing the SCIP ('Stellar, Circumstellar and Interstellar Physics') northern Galactic plane programme -- a strand within the overall WEAVE 5-year survey.  The software is also being deployed to find targets for the southern plane (based on VPHAS$+$ data), to enable similar observations via 4MOST.  In the text below, the acronym WEAVE appearing on its own will stand both for the instrument and for the WEAVE/SCIP survey strand, according to context. 

\subsection{Building lists of science targets with {\sc HaGrid}} 
\label{sec:hagrid2}

%To illustrate our methods and the outcome, we describe the construction of fields of WEAVE candidate targets.
A WEAVE pointing is defined by the coordinates of its centre and a field of view of radius of 1 degree, projecting a circle on the sky covering $\pi$ sq.deg.   Within such a field, the aim is to identify several hundred positions that coincide with regions of (locally) maximum H$\alpha$ brightness. 
%The process also seeks to return a list of positions that coincide with minimal brightness suitable for use as sky references.  % In both cases it is necessary also to screen out point sources, and identify exclusion zones around bright stars that may lie in the field.  

In outline, the steps taken are as follows:
\begin{itemize}
    \item Find all $H\alpha$ CCD images from the IGAPS repository in the area of interest.
    \item Mask out stars, CCD borders, bad pixels and vignetted areas from the data.
    \item Divide the image into superpixels.
%    \item Select the superpixels with the lowest counts as sky positions.
    \item Select the superpixels with the highest counts as candidate source positions.
\end{itemize}

The application of the algorithm, confronted with real data, is necessarily more complicated, particularly as it must deal as far as possible with all the artefacts that mimic real nebulosity.  So we now itemize the steps involved in more detail:
\begin{itemize}
    \item Collect the CCD images from the IGAPS repository in the area of interest.
    %This is a version of the same task as carried out by the repository search tool described in section~\ref{sec:website}.
    For one WEAVE field of radius 1 degree, this will be a list of over 100 H$\alpha$ CCD frames and their associated $r_I$ exposures.  Since frames with poor data quality often lead to false detections of H$\alpha$ emission, 
    %the following CCD images are 
    we excluded, where possible,
    grade C and D frames and favoured uniformly-calibrated over pipeline-calibrated data.
    %that did not enter the uniform IGAPS calibration; any frame obtained closer to the moon than 40 degrees (prompting bright sky background) not in the uniform calibration; grade C or D frames with seeing $<0.5$ arcsec (images at risk of high noise levels); %not calibrated data for which a calibrated alternative exists; 
    %frames named in a 'bad data' list generated by hand as outcomes were reviewed, which mostly contains images that show additional electronic noise or gain changes during readout.  Also, where the database holds both uniformly-calibrated and pipeline-only calibrated frames for a particular field ID, just the former are used.
    \item Create a star mask for each CCD image from both the IGAPS point-source database and a bright star catalogue to identify stars needing larger exclusion zones.
    The mask radius of IGAPS sources is set as a function of $r_I$ magnitude, H$\alpha$ seeing and ellipticity.
    %Stars above a certain brightness show diffraction spikes, which are also masked out. %(see left panel of figure \ref{fig:masking}).
    %For very bright stars (< 4.5 mag) a fixed size halo becomes visible in the data and, accordingly this is masked out (see right panel of figure \ref{fig:haloes}). The position of the halo relative to the star depends on the angular separation of the star from the optical axis of the telescope. %This is accounted for.
    Also masked are diffraction spikes of bright stars and a visible halo for very bright stars (< 4.5 mag, see right panel of figure \ref{fig:haloes}). The halo position relative to the star depends on the angular separation of the star from the optical axis of the telescope.
    
%    \begin{figure}
%    \centering
%    \colorbox{black}{
%    \includegraphics[width=0.45\columnwidth]{figures/appendix/r431452-1_mask.png}
%    }
%    \includegraphics[width=0.45\columnwidth]{figures/appendix/r541554-3_sel.png}
%    \includegraphics[width=0.45\columnwidth]{figures/appendix/r705492-4_sel.png}
    
%    \caption{Left: Example of masking on image r431452, CCD\#1. Masked pixels are shown in black. Brighter stars also have the spikes masked. A masked satellite trail is visible on the left side of the image.
%    Right: Example of a gain change during CCD readout. Left unmitigated, sky positions (blue circles) would be picked where the mean count level is lower, while source positions (red circles) are picked in the higher count region. To prevent this, a rectangular mask is applied by {\sc HaGrid} to the lower part of the frame where the gain is elevated.}
%   \label{fig:masking}
%   \end{figure}
    
    %\item Mask other artefacts.
    %CCD borders and pixels are masked that fall below a specified threshold in the linked confidence map (see \ref{app:confidence_map}). %This is constructed by the pipeline from flat-field information in order to identify bad pixels and vignetted areas. 
    %Hot or cold pixels are also masked that are found by hand, either via retrospective inspection of outstanding peaks in the distribution of (x, y) CCD pixel values linked to H$\alpha$ excess, or in CCD cutouts of selected targets. Artificial linear structures (satellite trails, noise bands, gain-change strips, bright star reflections, \dots) also found by visual inspection are masked.
    \item Other artefacts like CCD borders, pixels that fall below a specified threshold in the linked confidence map (see \ref{app:confidence_map}), hot or cold pixels and artificial linear structures (satellite trails, noise bands, gain-change strips, bright star reflections, \dots) found by visual inspection, are also masked.
    %Each one is defined by a start and end position, and a width.  %Figure~\ref{fig:masking} provides examples. 
    %\item Apply the mask and rank superpixels by $H\alpha$  brightness.  The relevant total mask is applied to each CCD frame, and the frame's area is divided into 'superpixels' to be appraised for contained counts.  A superpixel is a square of $n \times n$ native pixels, where $n$ is an adjustable input parameter.  %No superpixel extends beyond the frame edge.
    \item Create superpixels and rank them by \Ha brightness. Each masked CCD frame is divided into superpixels, which are squares of $n \times n$ native pixels, where $n$ is an adjustable input parameter.
    For WEAVE, $n = 25$ giving $8.25\times8.25$ arcsec$^2$ superpixels.  This choice tensions between good-enough angular resolution and the typical loss of area and statistics inflicted by the masking.
    Superpixels that are more than 50\% masked are rejected.
    %To prevent high counts due to particle hits from being mistaken for astronomical signal, the data are median filtered using 3$\times$3 pixel (about 1 arcsec$^2$) binning, before the mean count and variance within each superpixel is determined from the unmasked pixels. The superpixels are then ranked by mean count.
    To avoid particle hits being mistaken for astronomical signal, the data are median filtered using 3$\times$3 pixel (about 1 arcsec$^2$) binning. The superpixels are then ranked by mean count determined from the unmasked pixels.

%    \item Select sky positions.  For WEAVE, 11 (or fewer) sky positions with a minimum separation of 4 arcmin are found from the superpixel list. The search starts at the lowest mean count and ends when the maximum defined number of sky positions is found. A superpixel is rejected if it is closer than the distance limit to an already selected superpixel or if the variance exceeds twice its mean (suggesting e.g. a cold pixel or an instance of CCD cross-talk). The minimum of the 3$\times$3 pixel median-filtered data within the superpixel is found and adopted as the candidate sky position, while the sky value and its variance are set to the superpixel mean and variance. 
    %This guards against the sky value being influenced by cold pixels or CCD crosstalk, ie. negative images of bright stars on other CCDs.
%    The 25th percentile count level among the set of sky values is chosen as the sky value characteristic of the frame.
    %and the sky variance calculated. {\bf JED - previous sentence ??}

    \begin{figure}
    \centering
    \includegraphics[width=0.45\columnwidth]{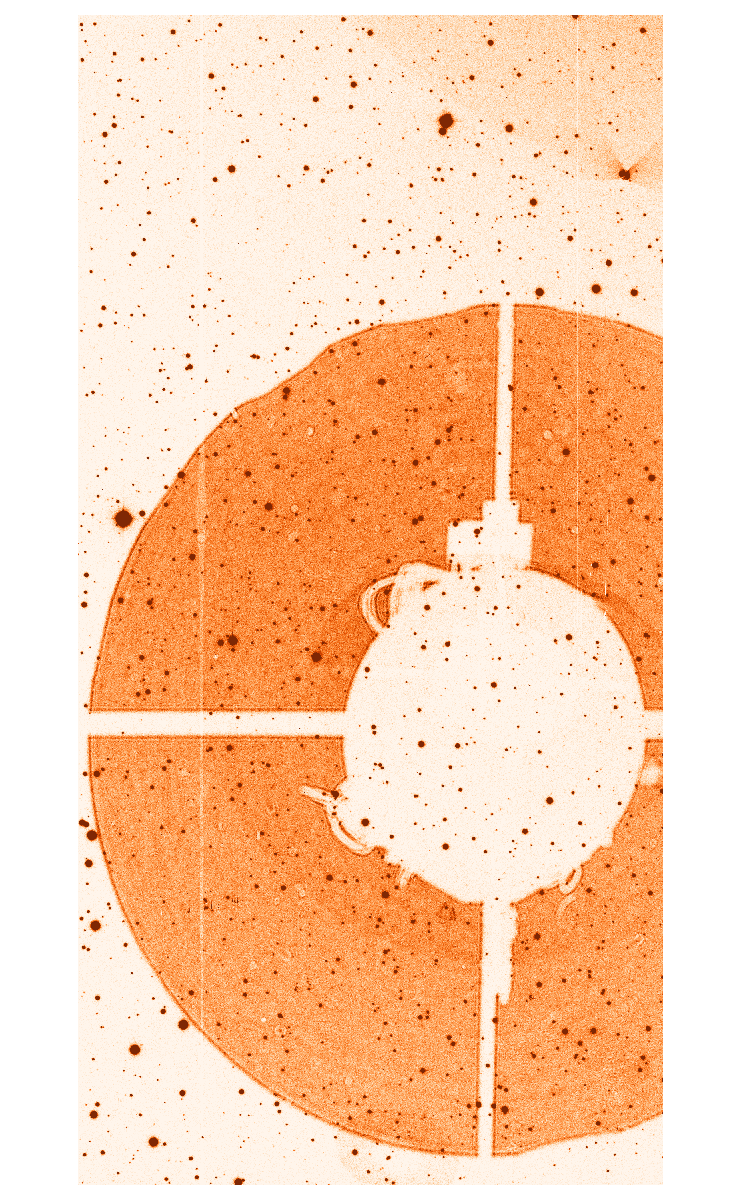}
    \includegraphics[width=0.45\columnwidth]{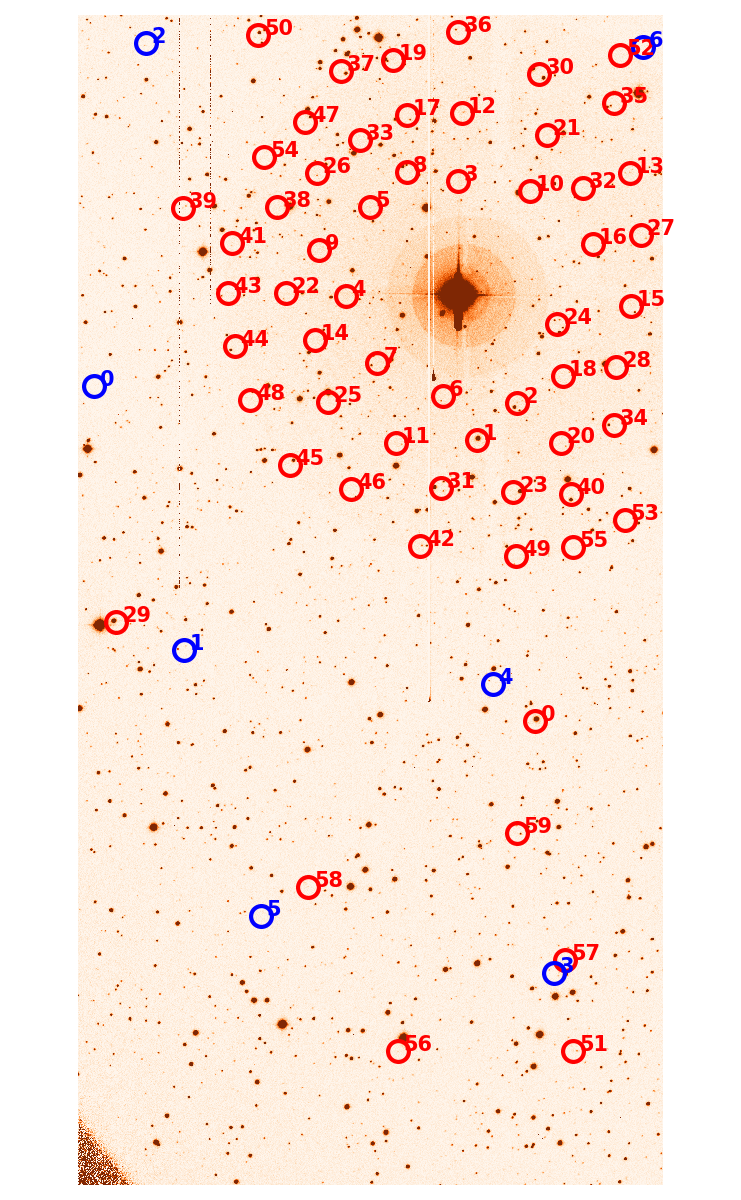}

     \caption{Left: Ghost on image r372056, CCD\#4 due to a bright star. Notice the stray-light image even picks up details of protruding cabling at prime focus of the INT.  Part of a much fainter ghost can be seen top right.
    Right: Example of a halo around a bright star fooling the {\sc HaGrid} selection from r541554 CCD\#3. Source positions selected by {\sc HaGrid} are shown as red numbered circles. They crowd into the faint extended halo that is a little offset from the star position. A larger stellar mask for bright stars and the rejection of 'star-like' selections as described in Section~\ref{sec:post_process} helps eliminate these. The blue circles mark identified counts minima that are stored as potential sky fibre positions.}
    \label{fig:haloes}
    \end{figure}

    \item Estimate the H$\alpha$ sky background.  A sound determination of the local sky value is very important, especially for determining the correct H$\alpha$ surface brightness. 
    %In areas of extensive and intense nebulosity, the algorithm so far measures the sum of sky and a significant astronomical background from the H$\alpha$ frames. In such areas, we also derive an estimate of the H$\alpha$ sky-only background value from the $r_I$ sky background value (calculated by the same procedure and for the same positions as for the narrow band). This uses a linear fit to a global plot of the $H\alpha$ sky level against sky level for $r_I$, exploiting the fact that most of the Galactic plane is free from nebular emission.
    The algorithm measures the sum of sky and any significant astronomical background from the H$\alpha$ frames. In predefined areas of extensive and intense nebulosity we also derive an estimate of the H$\alpha$ sky-only background value from the $r_I$ sky value. This uses a linear fit to a global plot of the $H\alpha$ against the $r_I$  sky level, exploiting the fact that most of the Galactic plane is free from nebular emission.
    %For each H$\alpha$ sky position the corresponding $r_I$ 3$\times$3 pixel mean filtered value is found as well, using the WCS to transform pixel coordinates between the different filters. 
    If the sky value inferred from $r_I$ is lower than the $H\alpha$ sky value, then we adopt the average of the two. 
    %as the $H\alpha$ sky value. 
    Taking the average was precautionary against problems with the $r_I$ sky prediction due to changing moonlight reflections into the telescope from clouds and other structures such as the dome.
    \item Select H$\alpha$-excess source positions from the ranked superpixel list for every CCD frame.
    %Another pass through the list is made, enforcing a minimum on-sky separation between any two entries (1 arcmin for WEAVE). The superpixel list is searched starting at the highest mean count: a superpixel is rejected if it is closer than the distance limit to an already selected superpixel. A check is also run on the difference between the superpixel mean and the frame H$\alpha$ sky: in all cases, this difference must exceed the sky noise (one sigma) if it is to remain in the list.
    The superpixel list is searched starting at the highest mean count. A superpixel is rejected if: it is closer than a distance limit (1 arcmin for WEAVE) to an already selected superpixel;
    the difference between the superpixel mean and the frame H$\alpha$ sky is less than the sky noise (one sigma).
    %\footnote{For WEAVE selection, a further  tweak was found necessary to deal with absolutely weak excesses of under 15 ADU: the threshold imposed is doubled to twice the estimated sky noise.}  
    %The maximum 3$\times$3 pixel median filtered value within the superpixel is found and its value, variance and position is stored. If the value - H$\alpha$ sky $< 10$ ADU, the superixel is rejected.
    Finally, the maximum 3$\times$3 pixel mean-filtered count within the superpixel is located and its position is adopted as the candidate target position.
    %, while the mean count and its variance are now stored.  
    If the difference between this more localised mean and the H$\alpha$ sky is $> 10$ ADU, the superpixel goes forward into a merged overview table. 
    \item The location, count, surface brightness and other data for each selected high-H$\alpha$ candidate position is appended to the overview table covering a large user-defined sky area, ready for further checking and analysis.  
    %A separate table of sky positions is also built.
\end{itemize}

As it searches for positions of bright H$\alpha$ emission, \hagrid also identifies suitable low-count sky positions and gathers statistics on sky noise. The distribution of sky noise versus sky background it finds closely resembles the distribution found by the pipeline shown in Figure~\ref{fig:sky_noise}.  %It is a bit lower because the \hagrid approach does not involve a fit over the whole CCD, and so avoids contributions from fainter stars.
The \hagrid found sky noise is lower as it does not involve a fit over the whole CCD, and so avoids contributions from fainter stars.
%where apart from sky also nebulosity and faint stars contribute to the determined sky noise, 
%the determined noise level is accordingly lower.

%A fit to equation \ref{eq:noise} as described in section \ref{sec:ha_properties} leads to an effective pixel number of $n_{pe}=201.5$ (to be compared with 2.88 \textbf{?? ...credible?? Robert A: well, its what the data says. The ideal value would be $25*25*\sqrt{11}=2073$}).

\begin{figure}
    \centering
    \includegraphics[width=0.85\columnwidth]{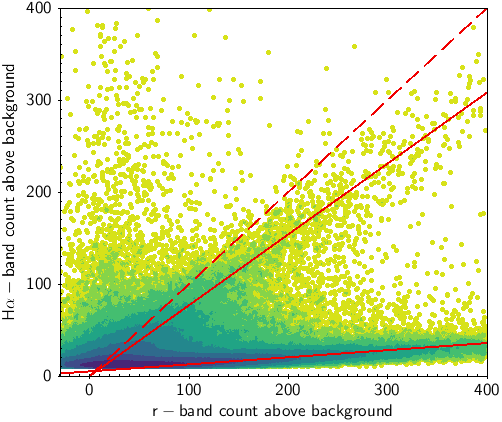}
    \caption{Excess $H\alpha$-band counts compared with the excess counts in the $r_I$ band.  Both quantities are the difference between the measured peak count and the estimated background level returned by {\sc HaGrid}.  Note that the \hagrid-output list is restricted to excess narrowband counts $>10$.  The data are plotted as a density map and before any second-stage cleaning.  The selection shown is from the sky region: ($30^{\circ} < \ell < 95^{\circ}$, $|b| < 4^{\circ}$). Working from top to bottom the red lines are: equality between counts (shown dashed), a line of slope 0.77 (solid) representative of nebula-dominated positions, and a line of slope 1/13 (solid) characteristic of star-like positions.}
    \label{fig:compare_counts}
\end{figure}

The algorithm can be applied to areas of arbitrary size. %For the final WEAVE catalogue we have run it on 1404 degrees$^2$, to cover a slightly over-sized version of the WEAVE footprint. 
As each CCD is independent of the others, the code parallelizes very effectively.
%For example, the complete WEAVE/SCIP footprint (46304 IGAPS CCDs) processes in under an hour when using 400 CPUs in parallel.
The choices of minimum distance between 
%sky and 
accepted source positions is driven mainly by the design of the destination wide-field spectrograph, and the anticipated observing strategy.  For WEAVE, applying the 1 arcmin minimum distance between fibre placements leads to at most $\sim$200 source positions selected per WFC CCD (with a much lower median of 5).

\subsection{Final processing: list cleaning and reduction} 
\label{sec:post_process}

\begin{figure}
    \centering
    \includegraphics[width=0.8\columnwidth]{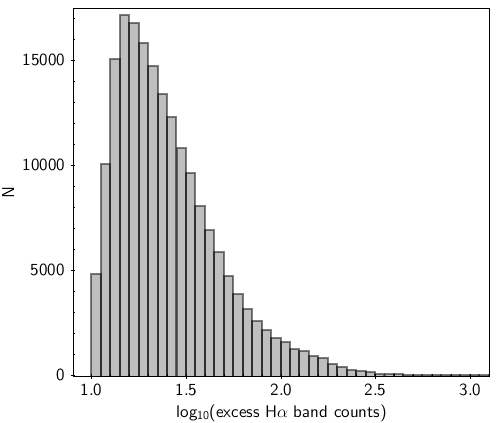}
    \caption{Final distribution of selected diffuse ISM sky positions for the WEAVE footprint as a function of the logarithm of excess $H\alpha$ counts.  It is highly skewed to low excess counts. A rough translation into surface brightness is that 10 excess counts, the minimum accepted, corresponds to $\sim3 \times 10^{-16}$ ergs cm$^{-2}$ s$^{-1}$ arcsec$^{-2}$. 
    }
    \label{fig:sb_histogram}
\end{figure}

\begin{figure*}
    \centering
    \includegraphics[width=1.8 \columnwidth]{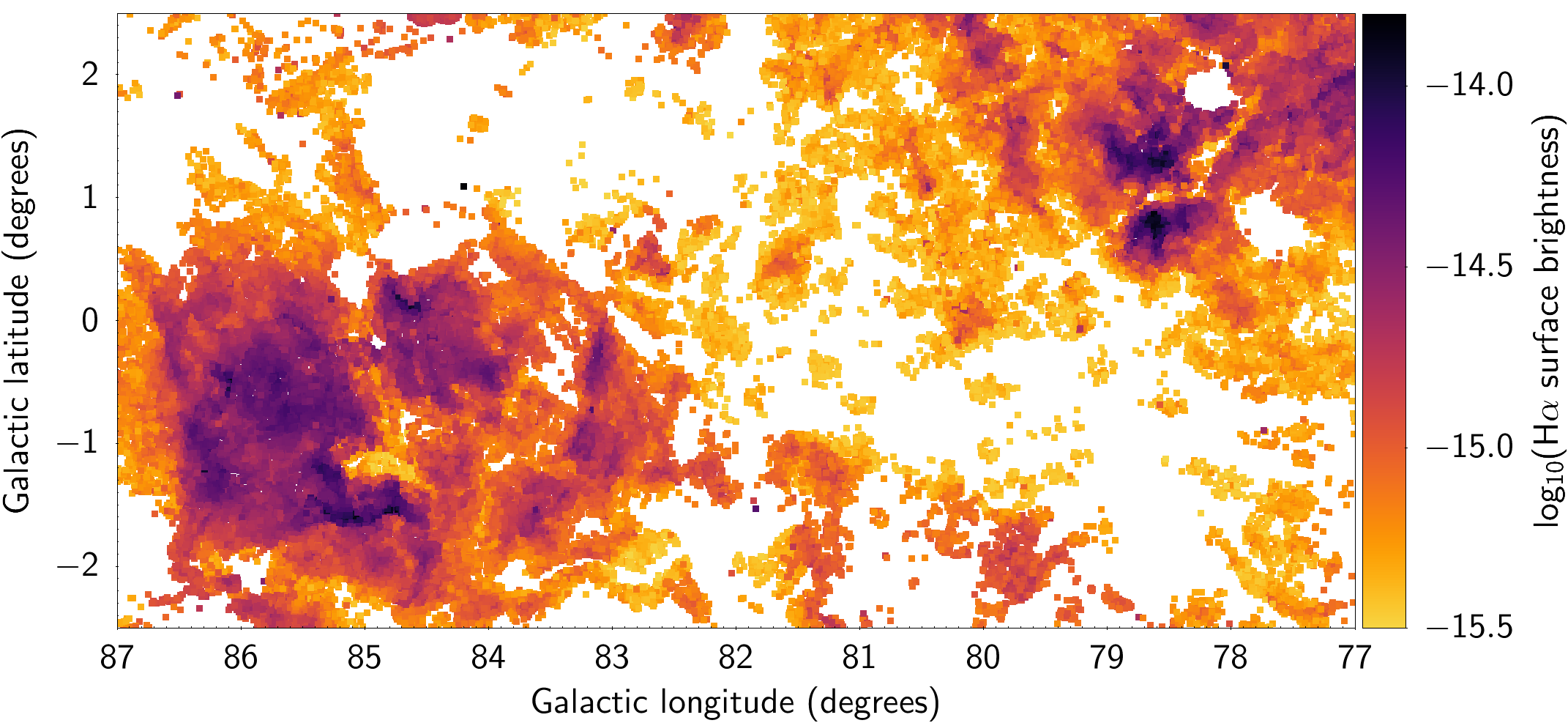}
    \includegraphics[width=1.8 \columnwidth]{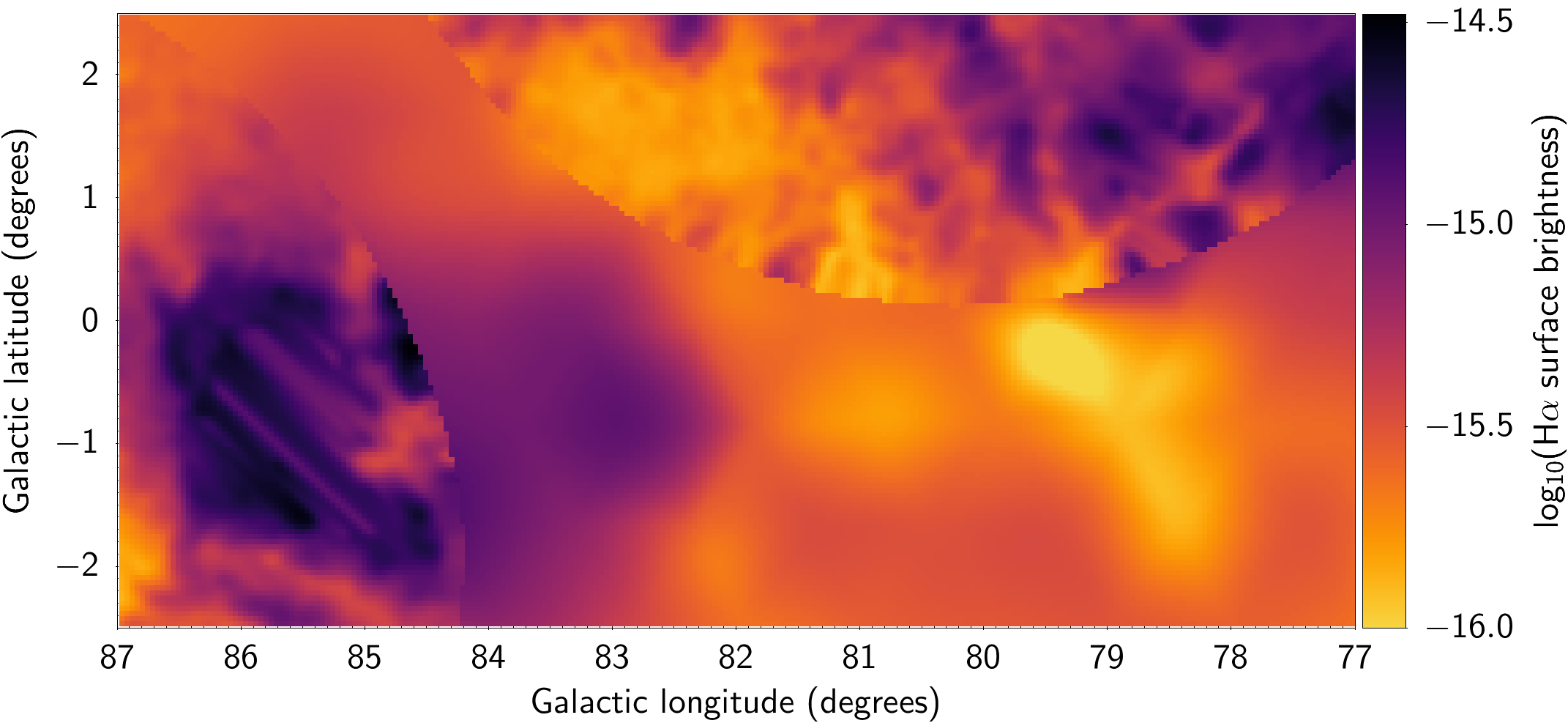}

    \caption{Top: Map of \hagrid\-selected positions in the Cygnus-X region of the northern Galactic plane. Every selected position has been coloured according to the logarithm of the estimated surface brightness (in ergs cm$^{-2}$ s$^{-1}$ arcsec$^{-2}$). White areas will have log(surface brightness) $< -15.5$. The region shown spans 50 square degrees or 3.6\% of the total area processed and contains 20\% of the identified candidate target positions.  It is the most line emission-rich part of the northern plane. 
    %The North America nebula can be made out to lower left (with the darker 'gulf' showing in yellow).
    Bottom: Same area in the \cite{Finkbeiner03} \Ha map, with 6~arcmin pixels.  The data available in this region is a combination of images from VTSS (circular footprint) and WHAM (making up the much coarser resolution background).
    }
    \label{fig:cygnusx}
\end{figure*}

The list of potential target positions generated as described above is long and needs further cleaning and reduction.  Not everything that appears bright in H$\alpha$ and is passed through by \hagrid has an astronomical origin: for example, some satellite trails, ghosts and unrecognised haloes around brighter stars can remain (Figure~\ref{fig:haloes}).  There are a number of further test-and-eliminate steps that can be taken to reduce the list to a high-confidence core. One such step is to favour repeat selection of the same emission structure and to reject isolated points (typically due to cosmic ray strikes that slip through). Generally speaking, we expect any given high surface brightness structure to be picked up twice or more, given that most sky locations are covered by a minimum of two images. 

An important piece of empirically-driven post-analysis is illustrated by Figure~\ref{fig:compare_counts} that compares the excess H$\alpha$ counts above the estimated background level with the excess obtained in the $r$ band, after scaling the latter to correct for the shorter exposure time.  Two main trends, drawn as solid lines, are apparent.  The uppermost of the two runs a bit below the equality line.  In the ideal case where H$\alpha$ and [NII] 654.8, 658.4~nm nebular line emission dominate the total counts measured in the $r$ band, the expectation would be that the measured count excesses in both the narrow and the broad band would be the same (given that the peak transmission in the narrow-band filter corrected for CCD response is closely comparable to the mean of the same quantity for the $r$ band).  This ideal does not apply, because of other nebular lines within the $r$ band, not captured by the narrow band (e.g. the [SII] 671.6, 673.1~nm doublet that can be strengthened by shock excitation, and potentially some [OI] 630.0, 636.2~nm emission). 
%Other lines of astronomical origin may also be present in emission (e.g. the [SII] 671.6, 673.1~nm doublet that is relatively strengthened by shock excitation).  
The approximate regression line shown in Figure~\ref{fig:compare_counts} has a slope less than 1, for this reason.  The objects of the search are indeed the candidate positions clustered around this empirical trend and, as such, they are the ones to keep.

In contrast, the second much lower gradient trend apparent in Figure~\ref{fig:compare_counts}, running close to the horizontal axis, is created by sky locations where the spectrum is continuum-dominated, i.e. star-like.  These locations can be stellar haloes where \hagrid picks up a seeming H$\alpha$ excess thanks to the typically wider seeing profile in the longer and unguided narrow band exposures, or ghosts (see Figure~\ref{fig:haloes} for examples).  In the case of a typical $0.5 \lesssim r - i \lesssim 1$ stellar continuum across the $r$ band, the expectation would be that the narrow-band excess counts would be approximately 1/13 of the $r$ counts -- this is the last of the three lines superimposed in Figure~\ref{fig:compare_counts}.  Candidate positions of this type need to be removed.   

%\textbf{JED ...paragraph needs more work.}
To make an accept/reject decision for every candidate position in the list, the distances to the expected nebular and stellar trend lines are calculated. These distances, $N$, are then expressed scaled to $\sigma$, the relevant Poisson-like error on the computed distance (subscript $n$ for nebular, $s$ for star-like).  This is followed by cuts applied in the $N_n, N_s$ plane to select the most credible nebular targets.  Inevitably, at low count levels, the confidence in assigning a candidate to the `nebular' and `star-like' classes weakens greatly.  The minimum excess count of 10 imposed by \hagrid helps deal with this, but a minimum cut on $N_s$ is also needed.  Where it is placed has to be tested empirically: for WEAVE we required $N_s > 3.5$.   

The selection can also be trimmed down to surface densities appropriate to the instrument used and the survey observing strategy (eg. number of visits, required science sampling). In the case of WEAVE this meant a 2 arcmin grid was placed over the relevant sky area and only two positions with the highest flux are kept in each grid cell.  Taking all the steps together for the WEAVE example, the original list of about 1.3 million target positions reduced to under 200\,000 potential targets, of which we expect around $1/4$ to be selected for observing.  

\subsection{Testing the down-size against known Herbig-Haro objects}

We have performed a retrospective test that compares the character of the long list with that of the final down-size by cross-matching them both with a list of known Herbig-Haro objects.  The latter has been established by a CDS criteria query using the term {\tt otype='HH'\&ra>0}. The coordinate condition was necessary to dismiss about 800 entries without coordinate information. The resulting table comprises 2622 positions: just 388 of them lie in the sky region defined by the Galactic coordinate ranges, $30^{\circ} < \ell  < 210^{\circ}$, $|b| < 4^{\circ}$ (roughly the footprint {\sc HaGrid} has been applied to).  When a limit of 10 arcsec is set on the angular separation, there are 70 and 40 successful cross-matches with the {\sc HaGrid} long and short lists, respectively. These numbers drop to 45 and 27 if the limit on angular separation is reduced to 5 arcsec. 

That no more than 20 percent, at best, of the listed HHO in the region are recovered is attributable to HHO position uncertainties, their high proper motion and the relatively low surface brightness of many. Another occasional factor at work will be the presence of substantial scattered starlight  lowering the contrast between the $H\alpha$ and $r$ images to below an acceptance threshold (emission in the vicinity of \object{V645~Cyg} is subject to this).  The most relevant point is that the down-sized list of candidate emission line positions captures more than half the number matching with the long list, despite the fact it contains only $\sim$0.15 as many positions. Proportionately, the shorter list is doing appreciably better, indicating that the downsizing has the side benefit of raising list quality. 
%The data contained includes measures of background counts, along with imported photometric zeropoints that provide the means to convert counts into surface brightness measures.  

\begin{figure}
    \centering
    \includegraphics[width=0.74\columnwidth]{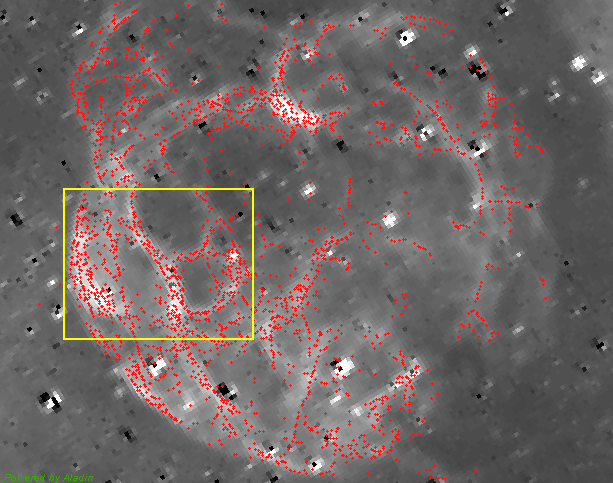}
    \includegraphics[width=0.74\columnwidth]{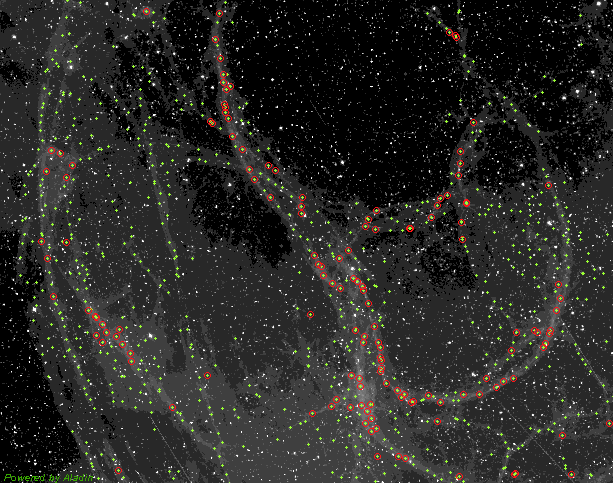}
    \caption{S147, the supernova remnant, as a test of {\sc HaGrid} position selection. The upper panel is a comparison between the pattern of final {\sc HaGrid} positions, in red, and the VTSS \citep{VTSS98} arbitrarily-scaled continuum-subtracted H$\alpha$ surface brightness map shown in greyscale. A good correspondence is achieved.  The lower panel is a zoom into a $0.27\times1.00$ sq.deg. area within the remnant superposing {\sc HaGrid} positions on the IGAPS $H\alpha$ mosaic at full $\sim$1 arcsec resolution (shown in full in Figure~\ref{fig:s147_mosaic}).   This shows how the fine filaments making up the remnant are accurately traced by the {\sc HaGrid} selections (green dots): those with surface brightness exceeding $10^{-15}$ erg cm$^{-2}$ s$^{-1}$ arcsec$^{-2}$ are encircled in red.  Both panels use Galactic coordinates such that latitude increases upwards and longitude towards the left. }
    \label{fig:S147}
\end{figure}

\subsection{Results of selection} 
\label{sec:hagrid_results}

The distribution of diffuse ionized emission in the northern plane is heavily weighted to low surface brightness and its presence along the plane is extremely uneven.  These are the outstanding features of the out turn from the application of \hagrid to the IGAPS images database.  The extent to which low surface brightness is favoured is illustrated by the histogram of excess $H\alpha$ counts, presented as Figure~\ref{fig:sb_histogram}.  The mean of the distribution is 32.5 excess counts, while the more informative median is just 21.6 (translating to a surface brightness of $\sim 7.2\times10^{-16}$ ergs cm$^{-2}$ s$^{-1}$ arcsec$^{-2}$ or close on 130 Rayleighs).

The upper panel of Figure~\ref{fig:cygnusx} is a cut-out of 50 square degrees in the Cygnus-X region of the plane, a region that is very rich in ionized gas: around 20 percent of all the selections are contained in just this 3.6 percent of the processed sky area.  Every data point is a candidate target position and is coloured according to the logarithm of H$\alpha$ narrowband surface brightness.  The figure amounts to a 'pointilliste' rendition of an $H\alpha$ image of the region emphasising the brightest locations. Comparison with the much coarser angular resolution H$\alpha$ map presented by \cite{Finkbeiner03}, in the lower panel of Figure~\ref{fig:cygnusx}, shows that the larger scale structure in H$\alpha$ surface brightness is shared, with the important difference that the {\sc HaGrid} output offers much more better-resolved, brighter detail that vanishes when smoothed to several arcminutes. The bright North America Nebula is located towards bottom left in both panels.  The hole in the emission in the top right corner of the upper panel is a product of needing to leave a zone clear around the 2nd magnitude star, $\gamma$~Cyg.  The larger hole below it is real in the sense that it is a dark cloud free of line emission.

A further example of the candidate positions output by \hagrid and the subsequent reduction is presented in Figure~\ref{fig:S147} for the case of S147 (already discussed and illustrated in Section~\ref{sec:mosaics}).  Its appearance is very filamentary, somewhat resembling a collection of soap bubbles, with the filaments at the interfaces.  The overall distribution of emission is captured very well by the selection, as shown by the comparison with the excerpt from the Virginia Tech Spectral-line Survey \citep[VTSS,][]{VTSS98}, in the upper panel of Figure~\ref{fig:S147}.  However the 1.6-arcmin pixel scale of VTSS does not entirely resolve the structure present. The filaments, with typical widths of under 1 arcmin, emerge more clearly in the IGAPS narrowband imagery, thanks to its native $\sim$1 arcsec angular resolution. This is picked up faithfully by the \hagrid selection.  The surface brightness of the selections ranges from $3 \times 10^{-16}$ to (very infrequently) $3 \times 10^{-15}$ ergs cm$^{-2}$ s$^{-1}$ arcsec$^{-2}$.

\section{Closing remarks}\label{Concl}

 A goal of this paper has been to present and describe the new IGAPS image database, formed from merging the data from both the IPHAS \citep{IPHAS} and UVEX \citep{UVEX} surveys of the northern Galactic Plane.  Around two-thirds of the database carries photometric zeropoints from the uniform calibration described previously by \cite{Monguio20}.  The collection is complete in that it contains all images of all qualities obtained over the course of the two long-running survey programmes.  This creates options to compare different epochs, given that many fields were observed more than once.
 
\begin{figure}
    \centering
    \includegraphics[width=0.8\columnwidth]{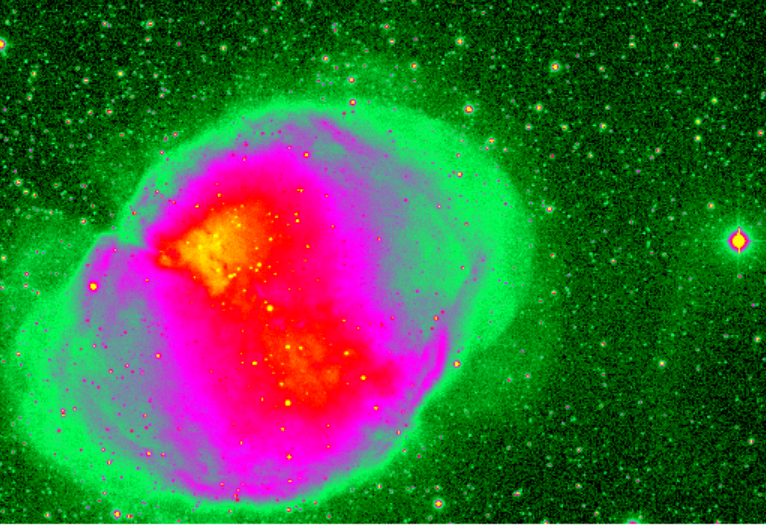}
    \includegraphics[width=0.8\columnwidth]{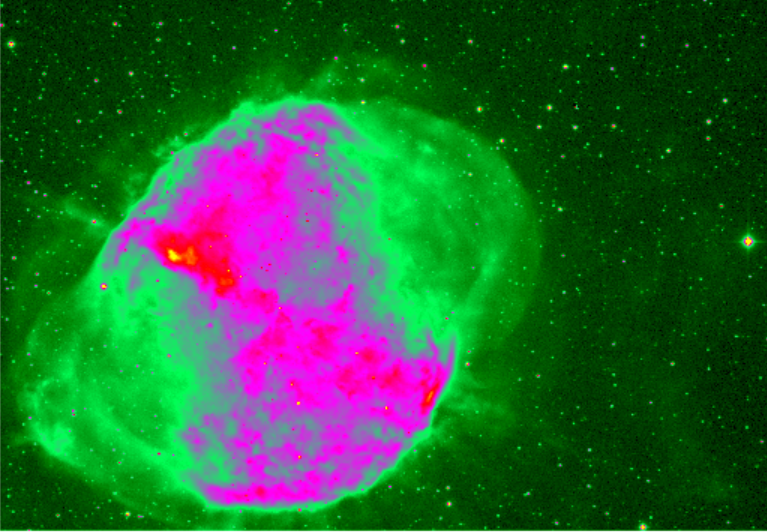}
    \caption{\object{Dumbbell Nebula}: as imaged in the $g$ and $H\alpha$ bands (upper and lower panels respectively).  The $g$ image includes the strong [O {\sc iii}] 495.9, 500.7 nm lines.  The cutouts are taken from the same corner of CCD \#2 in runs r-477762 and r-1241402, picked for having well-matched seeing (respectively 1.31 and 1.25 arcsecs).  The colour scales used are capped at 1250 counts for $g$ and 5000 counts for $H\alpha$ to match the ratio of their exposure times. They run from black at background level up through green/magenta/red to yellow at the bright end.  This PN has a diameter of approximately 6 arcmin. North up, east to the left. 
    }
    \label{fig:dumbbell}
\end{figure}

 The main focus of this paper has been on the $H\alpha$ narrow band:  the only one of the IGAPS set expressly targeting line emission.  Before summarising what we have presented on $H\alpha$, we recall that the Sloan $g$ band contains within its range the sometimes extremely bright [O {\sc iii}] 495.9, 500.7~nm doublet.  Accordingly, images taken using this filter can be used to compare and contrast the appearance of prominent nebulous regions in low ionization lines (H$\alpha$ and the [N {\sc ii}] 654.8, 658.4~nm doublet) and the much higher ionization [O {\sc iii}].  This will work especially well for lower extinction sightlines, where the [O {\sc iii}] lines are not disadvantaged by dust obscuration on top of the 1:4 exposure time ratio.  An outstanding example of such a comparison, for the Dumbbell Nebula, is shown as Figure~\ref{fig:dumbbell}. This is a large and bright planetary nebula in which it can be seen that the [O {\sc iii}] emission ($g$-band image, upper panel) is less clumpy than the H$\alpha$ + [N {\sc ii}] emission ($H\alpha$ image, lower panel), while the extent of the main nebula is nearly the same in both. Furthermore, even in the $g$ band 30-second exposure, fainter more extended structure is also apparent beyond the main nebula rim (seen to the west/right in the figure): its existence was first noticed in the lower resolution narrow-band images presented by \cite{Papama93}.  The central star also stands out at this shorter wavelength. The price paid for this kind of direct exploitation is, necessarily the strong pick-up of star light in the field because the $g$ band is broad.  Some improvement on this might be achieved by constructing $g - r$ difference images.

 Returning to our main aim -- we have provided a characterisation of the distinctive narrow-band $H\alpha$ data, whilst also showcasing some new illustrative applications.  Thanks to the \cite{Monguio20} zero points calibration, it is now more certain what the noise levels and sensitivities are: at full $\sim$1 arcsec angular resolution it is possible to distinguish nebulosity of surface brightness down to $\sim2 \times 10^{-16}$ ergs cm$^{-2}$ s$^{-1}$ arcsec$^{-2}$ (the typical noise level).  We have described here, in detail, how this may be exploited on the large scale to build target lists for diffuse-ISM spectroscopy using the coming generation of massive multiplex wide-field spectrographs (section~\ref{sec:hagrid} on the \hagrid algorithm).  On rebinning the native 0.333~arcsec pixels, the sensitivity increases as expected: \cite{Sabin2014} already claimed a typical sensitivity of $\sim 10^{-17}$ ergs cm$^{-2}$ s$^{-1}$ arcsec$^{-2}$ for 5 $\times$ 5 arcsec$^2$ binning.  We endorse this, with the necessary qualification that, in reality, there will always be a range in sensitivity, linked directly to the prevailing background level (see Figure~\ref{fig:sky_noise}).

 So far, most science exploitation of IGAPS image data has been directed towards planetary nebulae and, to a lesser extent, supernova remnants.  We anticipate this will continue into upcoming programmes using WEAVE and other new generation wide field spectroscopic instruments.  But there is also the opportunity to use especially the $H\alpha$ images to support science of the diffuse ISM in star-forming regions.  In particular the software tool, {\sc HaGrid} is permitting dense sampling across a wide area of many of the HII regions of the northern plane (as illustrated by Figure~\ref{fig:cygnusx}).  Target lists of this kind for multi-object fibre spectroscopy will be included in the WEAVE survey and should lead to new insights into the detailed chemistry and kinematics of the diffuse environment in and around young star clusters. A programme of a similar kind is already underway using LAMOST \citep{LAMOSTHa}: this samples the northern plane from a catalogue built also with the assistance of IPHAS images, now contained within the database presented here.
 
 The full collection of IGAPS images is available to the community via the website http://www.igapsimages.org.  This website provides an interface that facilitates downloads of
 selected images, served up as individual CCD frames. It also provides a number of related resources, including the \hagrid -generated list of bright diffuse H$\alpha$ northern-plane positions presented here.

\begin{acknowledgements}

This work is based on observations made with the Isaac Newton Telescope operated on the island of La Palma by the Isaac Newton Group of Telescopes in the Spanish Observatorio del Roque de los Muchachos of the Instituto de Astrof\'isica de Canarias.
This research has made use of the University of Hertfordshire high-performance  computing  facility  (https://uhhpc.herts.ac.uk/) located at the University of Hertfordshire (supported by STFC grants including ST/P000096/1).
This study has used part of an image obtained by the Virginia Tech Spectral-Line Survey, which is supported by the National Science Foundation. \\

We thank the following Bristol University students: Greg Mould, William Howie, Luke Davies, Heidi Naumann, Will Summers, Alex Townshend, Paul May, Matina Mitchell, Finn Hoolahan, Tom Burgess, Ashley Akerman, James Jordan, Simon Palmer, Anna Kovacevic, Jai Tailor, Olivia Smedley and Daniel Huggins for their work searching through difference images looking for PN candidates as part of their final year undergraduate projects.\\

RG benefitted from support via STFC grant ST/M001334/1 as a visitor to UCL.
JED \& MM acknowledge the support of research grants  funded  by  the  Science,  Technology  and  Facilities Council of the UK (STFC, grants ST/M001008/1 and ST/J001333/1).
MM was partially supported by the MINECO (Spanish Ministry of Economy) through grant ESP2016-80079-C2-1-R and RTI2018-095076-B-C21 (MINECO/FEDER, UE), and MDM-2014-0369 of ICCUB (Unidad de Excelencia 'Mar\'ia de Maeztu').   AM acknowledges support from the State Research Agency (AEI) of the Spanish Ministry of Science, Innovation and Universities (MCIU) and the European Regional Development Fund (FEDER) under grant AYA2017-83383-P.
PJG is partially supported by NRF-SARChI grant 111692 and acknowledges support from the Netherlands Organisation for Scientific Research (NWO), in contributing to the Isaac Newton Group of Telescopes and through grant 614.000.601.
\\

%This work has made use of data from the European Space Agency (ESA) mission
%{\it Gaia} (\url{https://www.cosmos.esa.int/gaia}), processed by the {\it Gaia}
%Data Processing and Analysis Consortium (DPAC,
%\url{https://www.cosmos.esa.int/web/gaia/dpac/consortium}). Funding for the DPAC
%has been provided by national institutions, in particular the institutions
%participating in the {\it Gaia} Multilateral Agreement.
Aspects of the analysis presented have been carried out via {\sc TopCat} and {\sc stilts} \citep{2006ASPC..351..666T}.
This research has made use of both the SIMBAD database and the "Aladin sky atlas", respectively operated and developed at CDS, Strasbourg, France.
% Here is the standard Montage acknowledgement (there isn't a reference last time I checked):
This research has also made use of the image manipulation software, {\sc Montage}. It is funded by the National Science Foundation under Grant Number ACI-1440620, and was previously funded by the National Aeronautics and Space Administration's Earth Science Technology Office, Computation Technologies Project, under Cooperative Agreement Number NCC5-626 between NASA and the California Institute of Technology.

%We thank the referee for comments on this paper that have improved its content.
\end{acknowledgements}

%   \style{aa} % style aa.bst
%   \{IGAPSAA} % your references Yourfile.bib

%\begin{the}{}
%   \bibliography{IGAPSAA} % your references Yourfile.bib
%\end{the}

\begin{appendix}
\section{On image properties and common artefacts}

%...compare and contrast cut outs of $\URGO$ and $g$ images to illustrate the different character of the U filter.  Go onto an outline discussion of image artefacts, including the vignetting commonly encountered in the corner of CCD3.
Here we comment briefly on the known $\URGO$ and $g$ PSF variations and present an example of a confidence map.  We then describe artefacts that one might come across in the survey imagery, especially in lower graded images. These are not unique to the WFC, as they can be found on most imaging instruments. We include this additional material to place it on record for the benefit of future users who may be less familiar with these oddities.
The survey data acquisition and pipelining has been described in previous papers \citep{IPHAS, IPHASIDR, UVEX, IPHASDR2, Monguio20}, with relevant aspects summarized here in section \ref{sec:images}. 

%{\bf THE FOLLOWING DOESN'T FLOW/MAKE SENSE IN CONTEXT - MOVE TO A.1?} Occasionally fringes can also be found at a very low level in $r$ frames. The effect is low enough that no defringing is applied. Residual $i$ frame fringes are usually visible at the $0.2\%$ level of the sky counts, though sometimes they can be more prominent (right image in figure~\ref{fig:odd_reflection}).

%\subsection{Image quality}
\subsection{$\URGO$ and $g$ PSF variations}
\label{app:image_quality}

\begin{figure*}
    \centering
    \includegraphics[width=0.95\columnwidth]{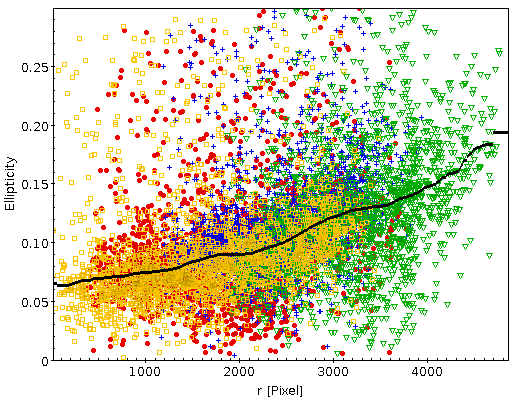}
    \includegraphics[width=0.95\columnwidth]{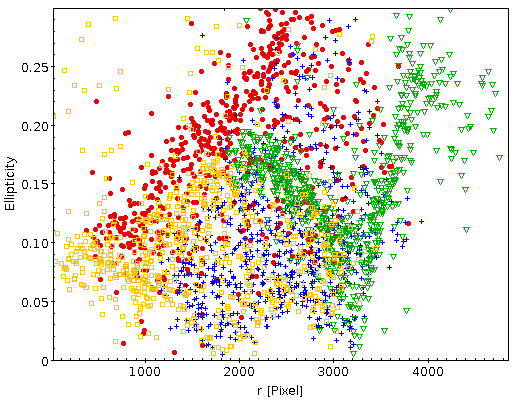}
    \caption{Dependence of pipeline measured source ellipticity on the distance to the instrument rotator center. The data is from IGAPS field 6395o, showing the sources from the $g$ image r584231 on the left, and from the $\URGO$ image r584230 on the right. On the left the data of all four CCDs is shown together with a running median plotted as a black line. On the right the data from each CCD is shown with different colours and symbols: CCD\#1 as red filled dots, CCD\#2 as blue plus signs, CCD\#3 as green triangles and CCD\#4 as orange squares.}
    \label{fig:ellipticity_WFC}
\end{figure*}

The $\URGO$ filter stands out among the filters used by the IGAPS survey in being a liquid filter: between 1mm UG2 and UBK7 glass plates lies a 5mm thick $CuSO_4$ solution. 

\begin{figure}
    \centering
    \includegraphics[width=0.95\columnwidth]{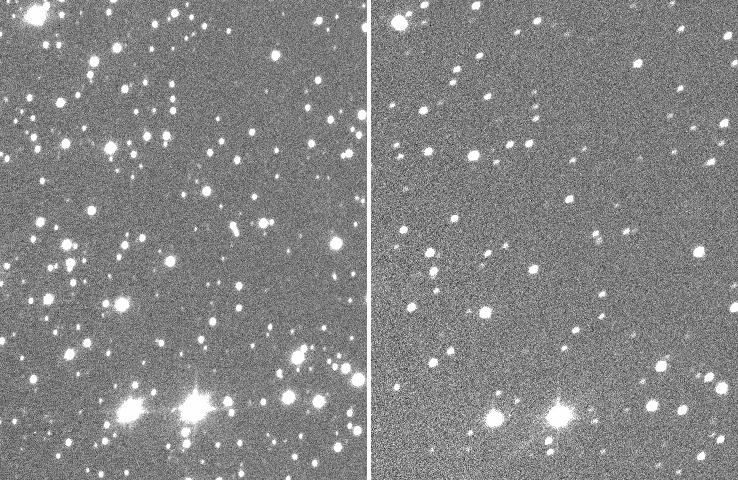}
    \caption{Cutout from CCD\#3 of images r584231 ($g$ filter, left) and r584230 ($\URGO$, right), showing the difference in PSF appearance between these filters.}
    \label{fig:RGO_U_cutout}
\end{figure}

The liquid nature of the filter leads to different image properties of the $\URGO$ data. Figure \ref{fig:ellipticity_WFC} shows a typical pipeline measured ellipticity distribution for the $g$ filter on the left, as an example representing the 4 normal glass filters used by IGAPS. The pixel distance to the instrument rotator center, which should be close to the optical axis, is shown on the x-axis. The data for all 4 WFC CCDs is combined in this plot. A running median is plotted as the thick black line. It can be seen that the ellipticity depends on the radial distance from the optical axis. The right panel of figure \ref{fig:ellipticity_WFC} shows the same information for the liquid $\URGO$ filter. A single radial trend does not exist. This leads to a more erratic point spread function (see the right-hand panel of figure \ref{fig:RGO_U_cutout}). It is also the reason why an additional 5th order term is sometimes necessary for the astrometric solution \citep{Monguio20}. It can also be seen in this example that, despite being a low extinction region, the number of sources visible in the 120 second $\URGO$ exposure is clearly less than in the 30 second $g$ exposure. The change in point source image morphology across the WFC array also has an effect on the morphological classification of sources, which are less likely to be classified as stellar the further they are located from the optical axis, independent of filter (cf. section 2.2 in \citealp{Farnhill2016}).

As the seal around the edges of the $\URGO$ filter is not completely tight, the contained solution slowly evaporates over time and hence needs to be topped up whenever an air bubble becomes apparent. Occasionally observations were made with a bubble visible in the filter, which always will drift to the filter edge in zenith direction. At minimum the red leak of the $\URGO$ filter would be increased for stars observed in the bubble area.  Accordingly, bright patches of stars near the edge of frames should be viewed with caution.

During the creation of the IGAPS catalogue it was discovered that there is also an optical blemish on the surface of the $g$ filter, that has an effect on the image quality. The location of this blemish changed over time, depending on the orientation in which the filter was reinserted into the filter holder after cleaning. For more information see section 6.1 and appendix B in \cite{Monguio20}.

\subsection{Confidence maps}
\label{app:confidence_map}

\begin{figure}
    \centering
    \includegraphics[width=0.95\columnwidth]{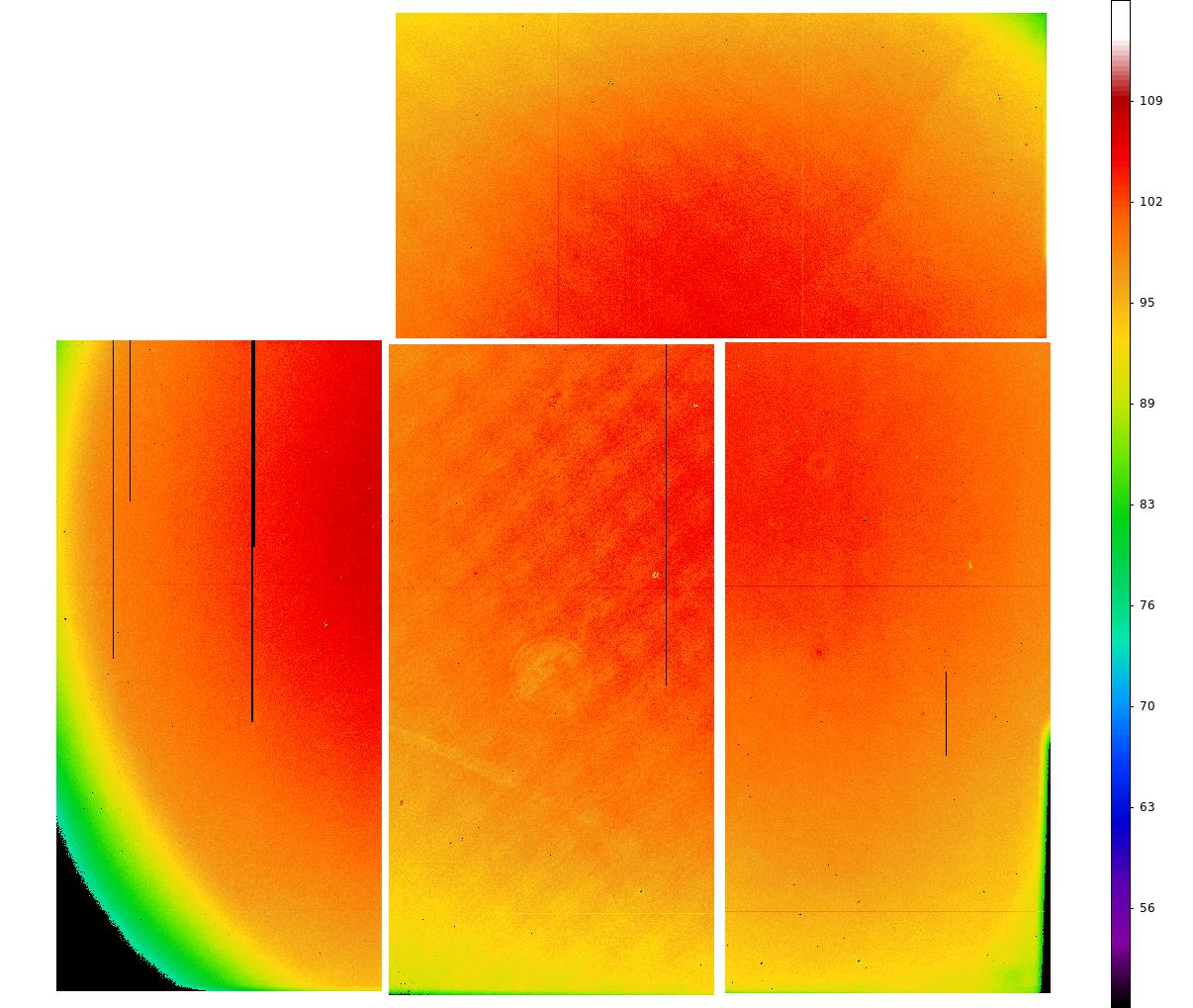}
    \caption{Confidence map for the \Ha filter from November 2012. Areas shown in black have a confidence value $< 50$. The layout of the CCDs has \#2 on top, and \#3, \#4 and \#1 from the left at the bottom.}
    \label{fig:confmap_WFC}
\end{figure}

The confidence maps produced in pipeline processing are used for the masking of bad pixels and vignetted areas on the CCDs. The pipeline produces the confidence maps per filter from the observations of flatfields taken during an observing run. Hence each IGAPS observing run has its own set of confidence maps associated with it. The confidence map is referenced in the FITS header item CONFMAP (see appendix \ref{app:header}).
To define bad pixels, often a limit of confidence $< 90$ is used. Figure \ref{fig:confmap_WFC}
shows an example of a confidence map. The vignetting of the image area is clearly visible.  The use of a round filter in the WFC has its biggest impact on CCD\#3, but also affects corners of CCD\#2 and CCD\#1. 

The worst column defects are also visible in the figure. Many thinner column defects and small bad pixel areas are not visible at this image resolution.
For specific purposes, such as the selection of diffuse-ISM targets for wide field spectrographs as described in section \ref{sec:hagrid}, the automatic selection of bad pixels via the confidence maps can benefit from the addition of further hot and cold pixels identified by the user.

\subsection{Artefacts}
\label{app:artefacts}
A number of artefacts can be found on the IGAPS WFC images, just as they are in data from a range of astronomical cameras.
%In general these are no different to artefacts found with other instruments, although the exact form and incidence rate will vary between instruments.
The cause of these are either optical reflections, external particle or light sources or electronic components.

\subsubsection{Bright stars}

For bright stars even the faintest parts of the point spread function become visible.
The right panel of Figure \ref{fig:haloes} shows a bright star and its associated halo, with a radius of $\sim$1100 pixels.
It can be seen that the halo is offset from the star. This offset depends on the distance to the optical axis of the WFC, which hints at the reflection being caused by a curved optical surface. As shown by this example, the halo of a bright star can affect a considerable part of a CCD.

Apart from the direct image and halo, the light of a star also gets reflected on different optical surfaces of the instrument and forms several large 
reflections, that can appear more than a degree from the star ('ghosts'). In the case that a star is very bright, these reflections will become visible in the images. One example is shown in the left panel of figure \ref{fig:haloes}. The reflection nearly covers a full CCD, and a lot of fine detail from the telescope entrance pupil can be made out -- including the cabling of the WFC at the prime focus that protrudes beyond the central obstruction. Also visible is part of a fainter large reflection in the top right corner of the CCD. The seeming small bipolar nebula at the bottom of that fainter reflection is not real either. That it is just reflected light can be checked by inspecting the offset partner image of the same sky location.

\begin{figure}
    \centering
    \colorbox{black}{
    \includegraphics[width=0.45\columnwidth]{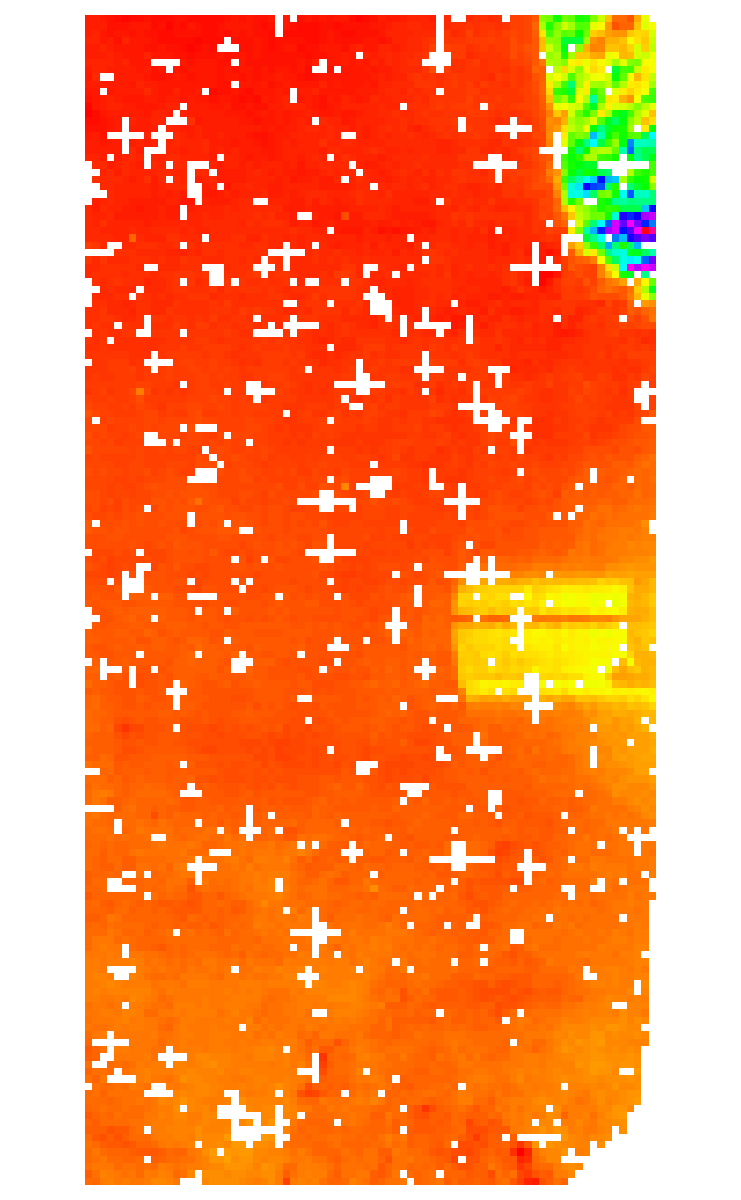}
    \includegraphics[width=0.45\columnwidth]{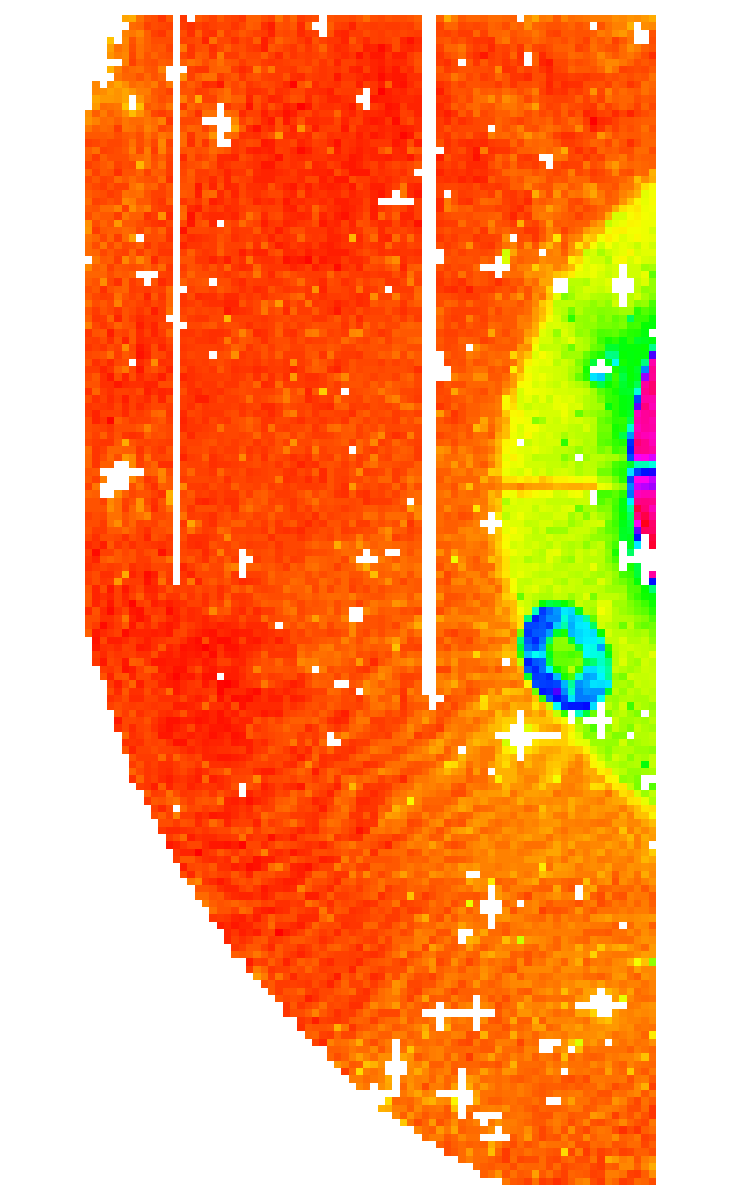}
    }
    \caption{Reflections of bright stars outside the CCD. Left: image r430532, CCD\#1, $H\alpha$ filter. The reflection in the center is up to 25\% of the background. The one near the top reaches more than 300\% of the background counts. Right: image r764550, CCD\#3, $i$ filter. The large and small reflection have 6\% and 21\% more counts than the background, respectively. The visible fringing is at about 1.5\% of background. The \hagrid superpixel map is shown to enhance low level detail. Black areas correspond to rejected superpixels due to high levels of masking. The color scale goes from red (low values) to blue (high values).}
    \label{fig:odd_reflection}
\end{figure}

Reflections do not always have to appear complete or have a circular shape.
Figure \ref{fig:odd_reflection} shows the superpixel map created by {\sc HaGrid}, as described in section \ref{sec:hagrid2}, as this makes low level detail more readily visible. The left panel of that figure shows two odd reflections. One with a square appearance, which is actually just a cutout from a much larger circular reflection. The other one visible in the top right corner of the CCD is quite different in appearance from the usual circular reflections. The right panel shows part of a circular reflection from a star outside the CCD. On top of it is a smaller, more elliptical reflection. Also visible in this superpixel map is low level fringing outside of the areas covered by the reflections. And the effect of vignetting on CCD\#3 is also clearly discernible.

\begin{figure}
    \centering
    \includegraphics[width=0.45\columnwidth]{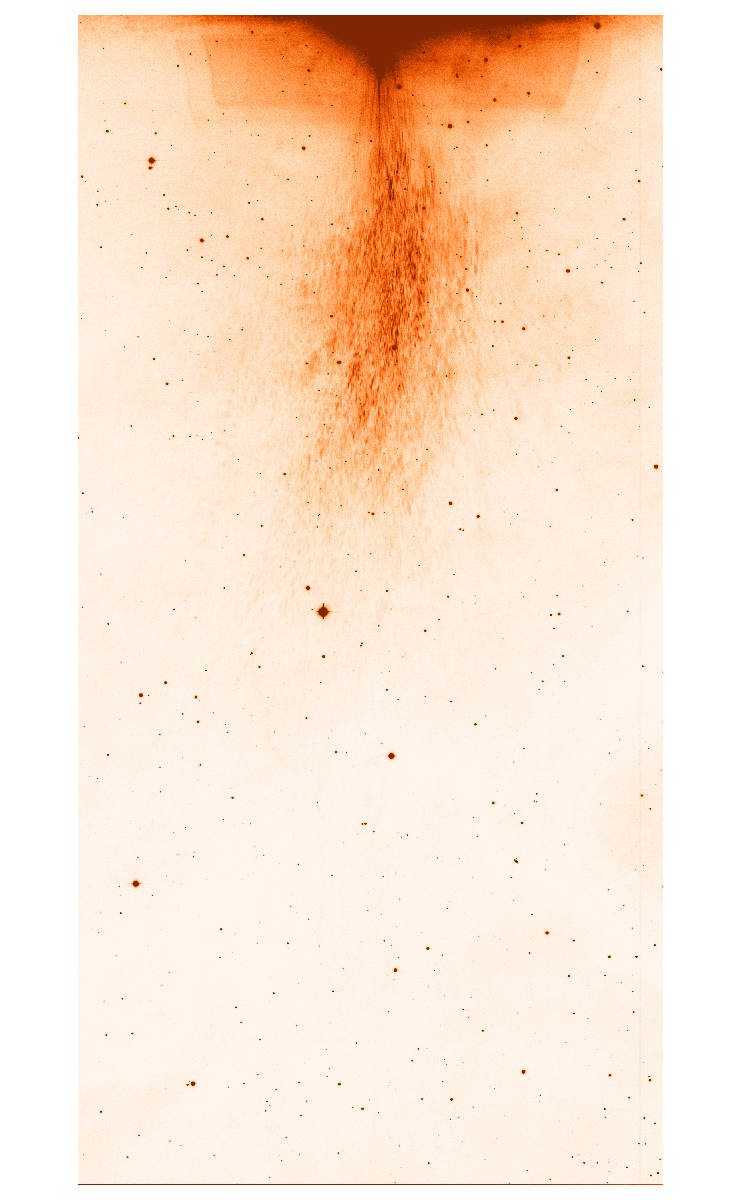}
    \colorbox{black}{
    \includegraphics[width=0.45\columnwidth]{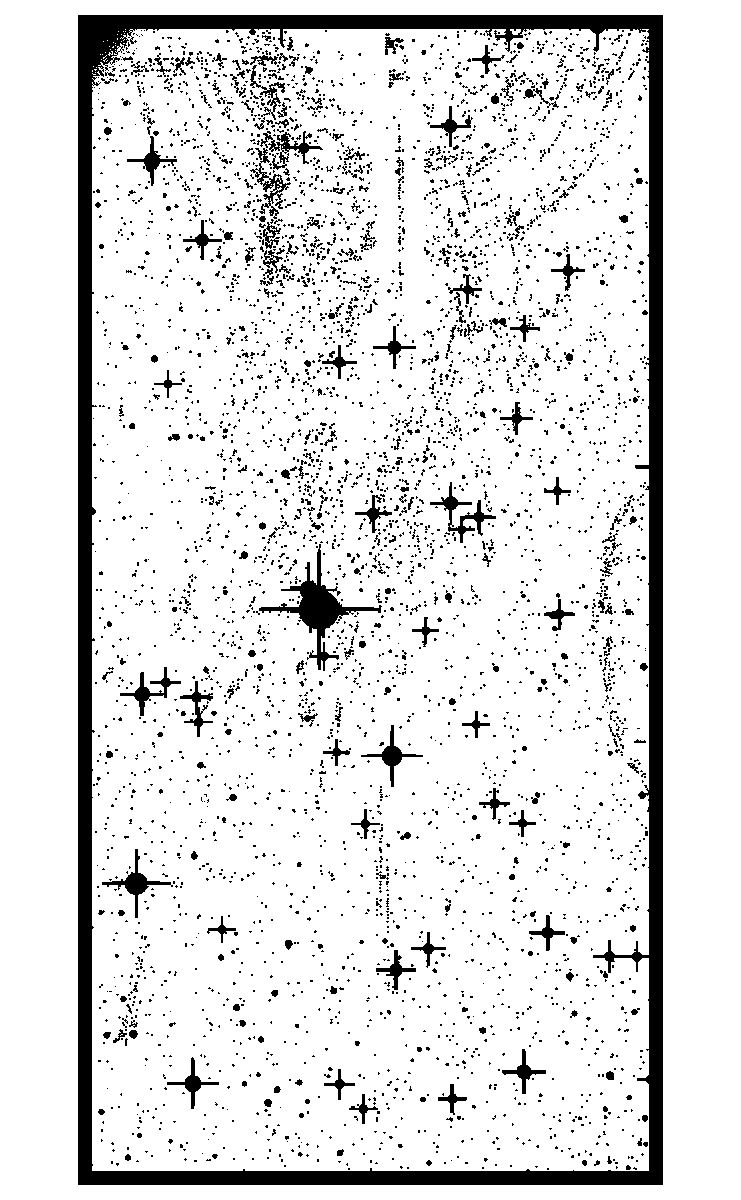}
    }
    \caption{Reflection of a bright star on image r414326, CCD\#2. On the left side the \Ha image is shown. The pixel mask is shown on the right. Black corresponds to masked pixels. The bright star is located just outside the top edge of the CCD. This creates a cometary like tail that covers about half the CCD length and a square reflection near the CCD top. The usual circular reflection is also visible at the top. As the masked pixels are generated from the catalogued sources it can be seen that a lot of faint spurious sources are detected due to the reflections.}
    \label{fig:bright_414}
\end{figure}

Figure \ref{fig:bright_414} shows the effect of a bright star located at or near the edge of a CCD. This leads to the starlight being reflected on to the CCD at an oblique angle. The left panel shows that this creates a cometary tail like structure, that can extend quite a distance from the star. Also an almost-square reflection is visible near the top edge, along with a fainter circular reflection. The mask created by \hagrid (cf. section \ref{sec:hagrid2}) is shown in the right panel.  It can be seen that this reflection creates a lot of faint spurious sources, picked up erroneously by the IGAPS catalogue, which are arranged in the circular pattern of the reflection.  Further spurious sources can be seen extending along the cometary reflection. A further faint part of a circular reflection must exist at the centre of the right CCD edge, as another circular structure of spurious sources can be seen there.

\subsubsection{Cosmic Rays and satellite trails}

\begin{figure}
    \centering
    \includegraphics[width=0.45\columnwidth]{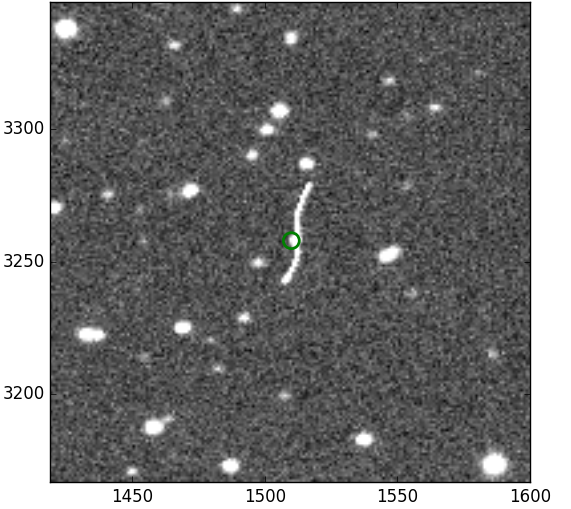}
    \includegraphics[width=0.45\columnwidth]{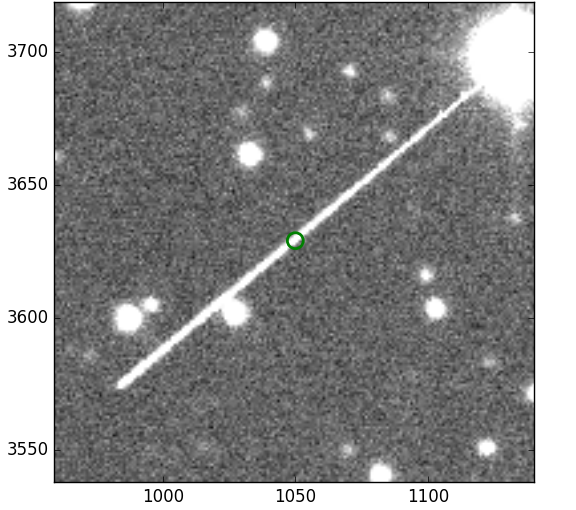}
    \includegraphics[width=0.45\columnwidth]{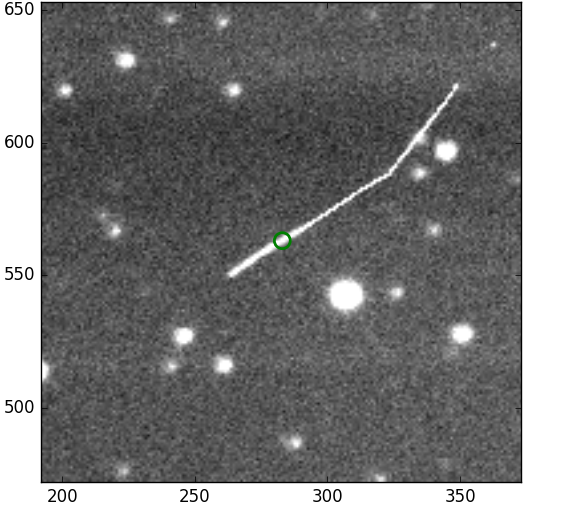}
    \includegraphics[width=0.45\columnwidth]{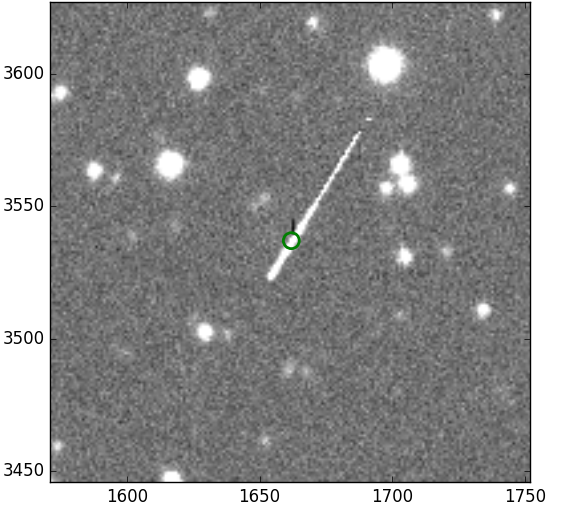}

    \caption{Examples of prominent cosmic ray impacts. Top left: A worm caused by multiply scattered low energy electrons. The other three panels show muon impacts. %The image on the bottom left shows a kinked track.
    The images are: top left, r372612 CCD\#4; top right, r418359 CCD\#2; bottom left, r431220 CCD\#2; bottom right, r570439 CCD\#3. Each cutout is 180 pixels squared.
    }
    \label{fig:cosmic}
\end{figure}

Cosmic ray impacts are a well known nuisance in astronomical images \citep{2002SPIE.4669..172S}. Despite their name, not all impacting particles actually are of direct cosmic origin. Figure \ref{fig:cosmic} shows a few prominent examples of particle impacts visible in the 120s \Ha exposures. The long streaks visible in the right hand panels are caused by muon impacts. The track in the lower left panel shows a kink in the track, probably due to a collision with a particle in the CCD. The top left panel shows a so called "worm", caused by multiply-scattered low energy electrons.

\begin{figure}
    \centering
    \includegraphics[width=0.46\columnwidth]{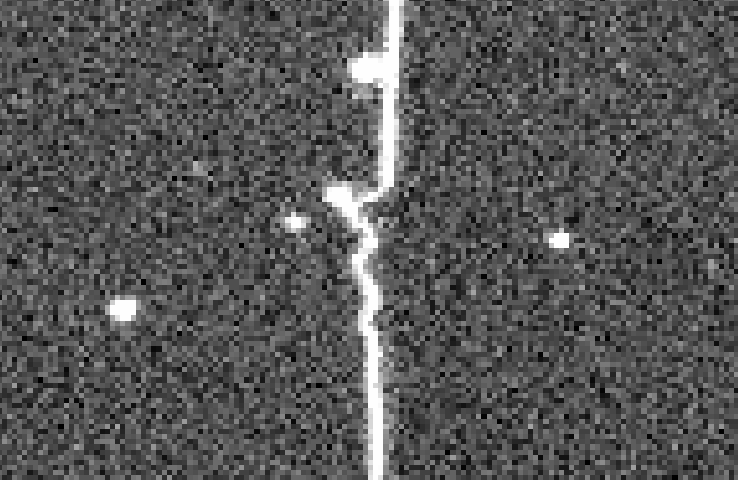}
    \includegraphics[width=0.46\columnwidth]{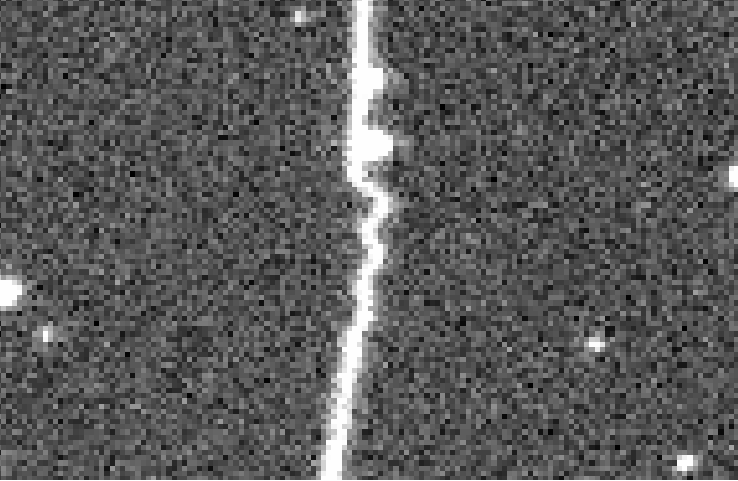}
    \caption{Details of the satellite track on the \Ha image r541904, CCD\#3. Two excursions away from linear are shown. Telescope tracking or satellite structure might have caused them.  }
    \label{fig:sat_structure}
\end{figure}

Another nuisance in astronomical images are satellite tracks. With the current and future planned mega satellite constellations in low earth orbit, this problem is very much on the increase. Satellite tracks are mostly straight lines running through the image at an angle. Sometimes flares can be seen, where the brightness increases for a short time due to the alignment of reflecting satellite surfaces with the observing direction. Rather rare is the observation of fine structure in the satellite track. One such example can be seen in figure \ref{fig:sat_structure}.
The cause of these high frequency "wiggles" may either be due to telescope-tracking glitches in declination or to satellite spin bringing different structures into illumination.
%Three excursions from the slightly curved satellite track, two to the left and one to the right, are visible in the image. Each of these excursions shows further fine structure, as can be seen from fitting a running gaussian to the satellite track. The seeing of this image was 0.7 arcsec (about 2 pixels), and the deviation of the excursion from the general track was up to 8 pixels. This means that the reflecting structure would measure between 2 to 12 meters in size if the object was at typical low earth orbital altitudes. The track also showed a flare of about 2 magnitudes. %As the track started (or ended) outside the CCD, only upper limits for the rotation period, orbital speed or flare duration would be possible to derive.

%\begin{figure}
%    \centering
%    \includegraphics[width=0.45\columnwidth]{figures/appendix/r431452-1_sel.png}
%    \colorbox{black}{
%    \includegraphics[width=0.45\columnwidth]{figures/appendix/r431452-1_mask.png}
%    }
%    \caption{A satellite track on image r431452, CCD\#1. On the left side the \Ha image is shown. The mask is %shown on the right, with masked pixels shown in black. The track shows a flare event.}
%    \label{fig:sat_431}
%\end{figure}

\subsubsection{Pickup noise and gain changes}

\begin{figure}
    \centering
    \includegraphics[width=0.45\columnwidth]{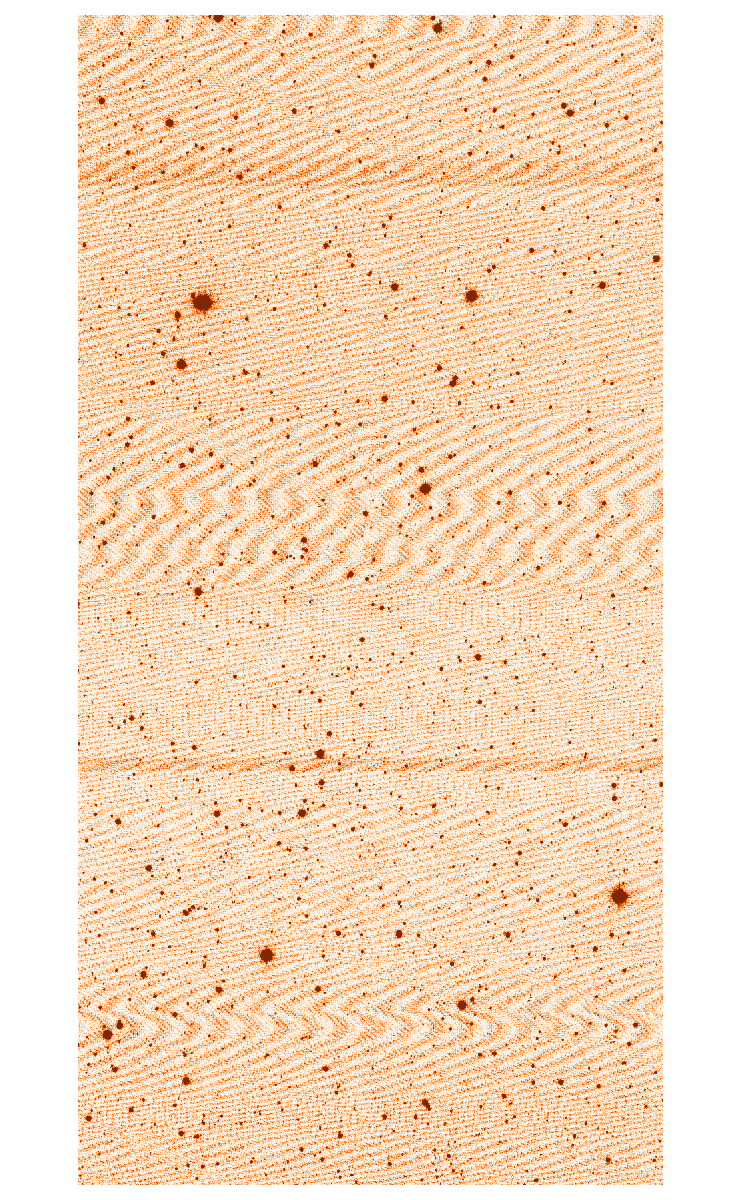}
    \includegraphics[width=0.45\columnwidth]{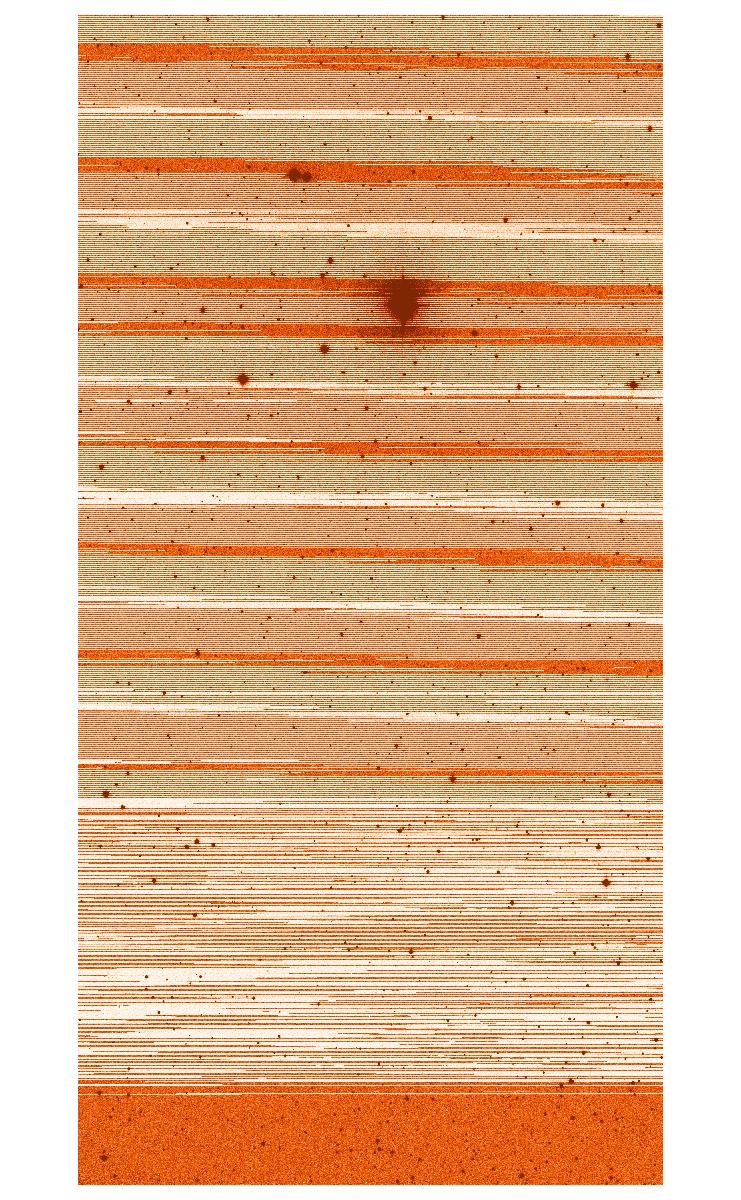}
    \caption{Left: Pickup noise on image r1018959, CCD\#4.  The variations here are small, and represent minor degradation.  Right: Gain changes on image r1166073, CCD\#4. The changes are both spread across the whole and large, rendering this frame unusable.}
    \label{fig:ccd_noise}
\end{figure}

Occasionally the WFC images suffer from electronic noise, either from external sources (pick up noise) or from readout electronic problems (gain changes). Over time, with ageing electronics, especially the gain changes became more common. The occurrence of gain changes seems to be random, and subsequent images usually are read out correctly. In most cases there is only one gain change during the readout. %as can be seen in the right hand panel of figure \ref{fig:masking}.
As certain observing runs had an increased occurrence rate of gain changes, not all of the affected fields could be re-observed at a later date. Hence the data reduction pipeline was modified to deal with the gain changes and still produce a useful object catalogue. An extreme example of many gain changes during the readout is shown in figure \ref{fig:ccd_noise} in the right panel. These extreme cases could not be salvaged by the data reduction.

The left panel of figure \ref{fig:ccd_noise} shows an image with typical pick up noise. 

\subsubsection{Cross talk}

\begin{figure}
    \centering
    \includegraphics[width=0.9\columnwidth]{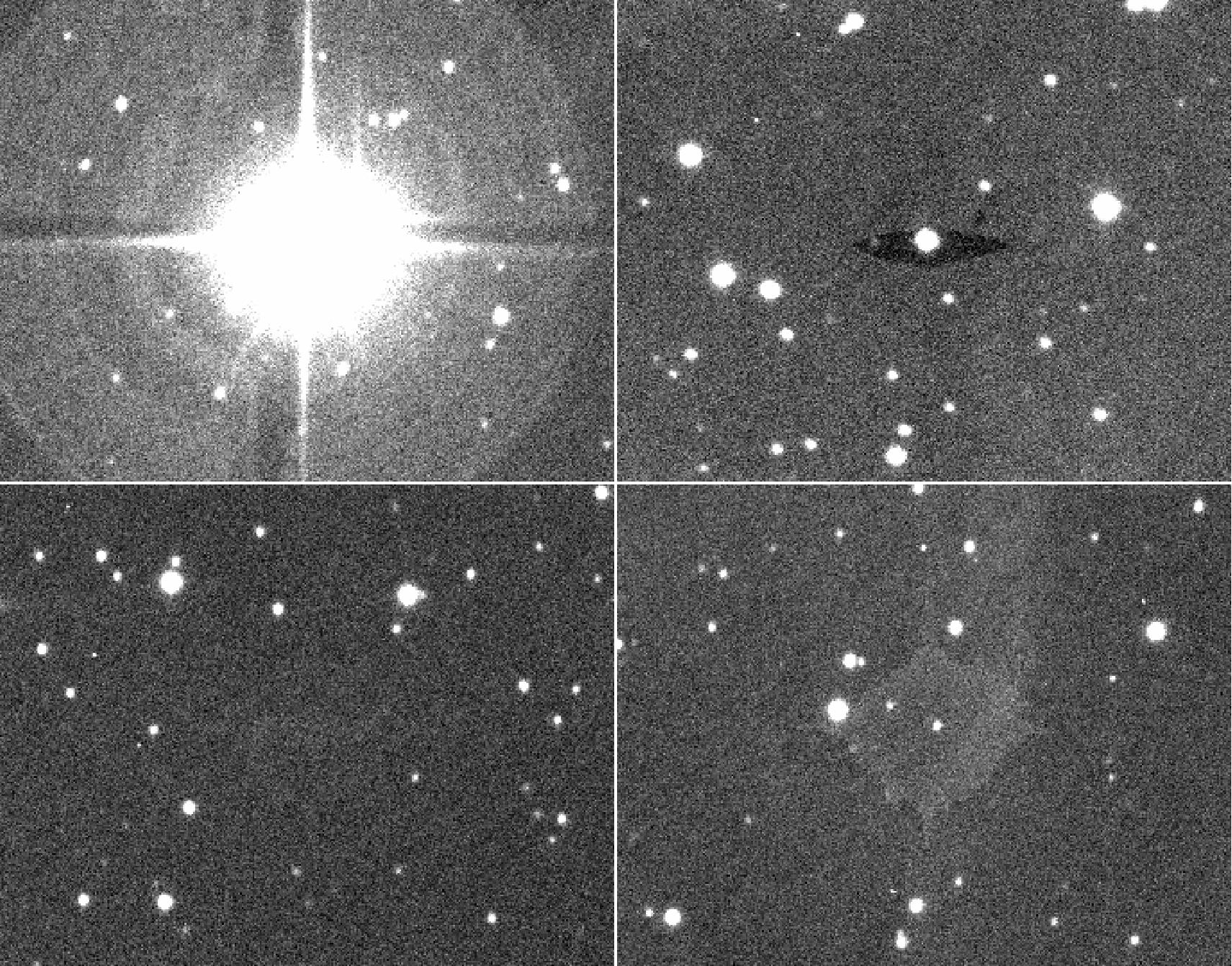}
    \caption{Cross talk on image r1023036. CCD\#1 is shown top left, \#2 top right, \#3 bottom left and \#4 bottom right. The same pixel section is shown for each CCD. The bright star on CCD\#1 is seen as a negative imprint on \#2 for pixels that are saturated, as a positive imprint on \#4 and a faint positive imprint on \#3.}
    \label{fig:crosstalk}
\end{figure}

Cross talk during readout is a well known phenomenon for CCD arrays \citep{2001ExA....12..147F}. With the WFC, cross talk only becomes visible when a very bright source falls on one of the CCDs.
An example is shown in figure \ref{fig:crosstalk}. The bright star in CCD\#1 creates a negative cross talk in CCD\#2 at a level of about -10 ADU. CCD\#4 shows a positive cross talk at about 10 ADU, and CCD\#3 only shows a very small positive cross talk signal in this case.

%\begin{figure}
%    \centering
%    \includegraphics[width=0.45\columnwidth]{figures/appendix/sky-r1023036-2-0.png}
%    \includegraphics[width=0.45\columnwidth]{figures/appendix/sky-r1170016-2-0.png}
%    \caption{Cross talk examples.}
%    \label{fig:crosstalk}
%\end{figure}

\subsubsection{Multiple images}
\begin{figure}
    \centering
    \includegraphics[width=0.9\columnwidth]{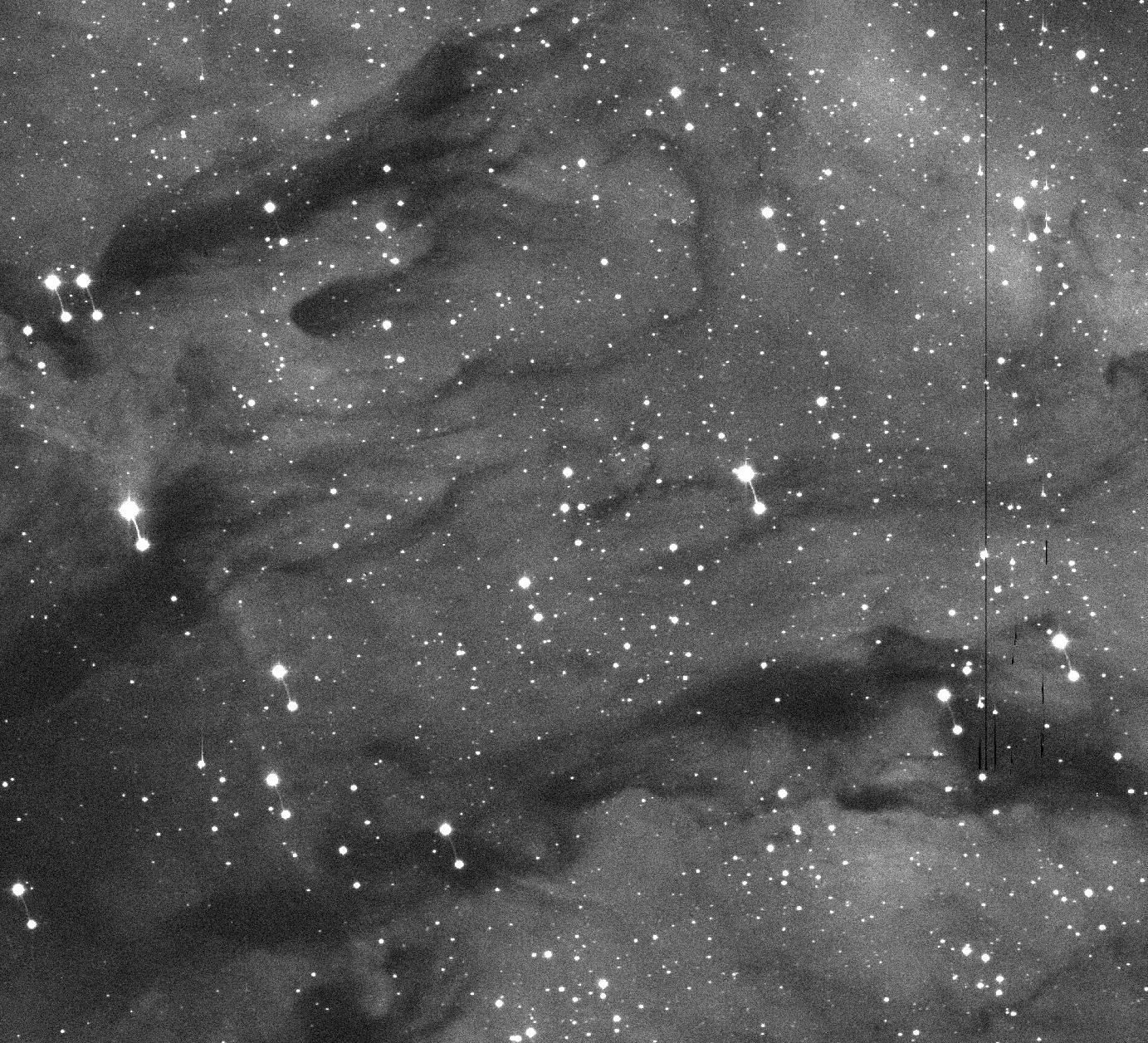}
    \caption{Double images on frame r367586, CCD\#4. Bright stars clearly show the sign of INT mirror support oscillation, forming dumbell like images.}
    \label{fig:double}
\end{figure}

Very rarely, multiple images of each source are found on an IGAPS exposure. Normally these appear as double images or streaks. The latter effect is caused by the telescope not being settled on the observing position by the time of exposure start. The former is caused by jumps in the telescope tracking or, in the case of the INT, the oscillation of the main mirror support system (see figure \ref{fig:double}). The mirror support system at the INT consists of 36 pneumatic pads, which are controlled together in three $120\deg$ segments. Oscillations of the servo loop were audible in the control room and could be stopped by the observer by moving the telescope to a different position. Note that due to differences in the time spent at the end points of the oscillation the double images are of different brightness.
% a description of the mirror support system can be found at
% http://www.ing.iac.es/~eng/electronics/int/telescope/int_tele.html

\subsubsection{Other}

\begin{figure}
    \centering
    \colorbox{black}{
    \includegraphics[width=0.45\columnwidth]{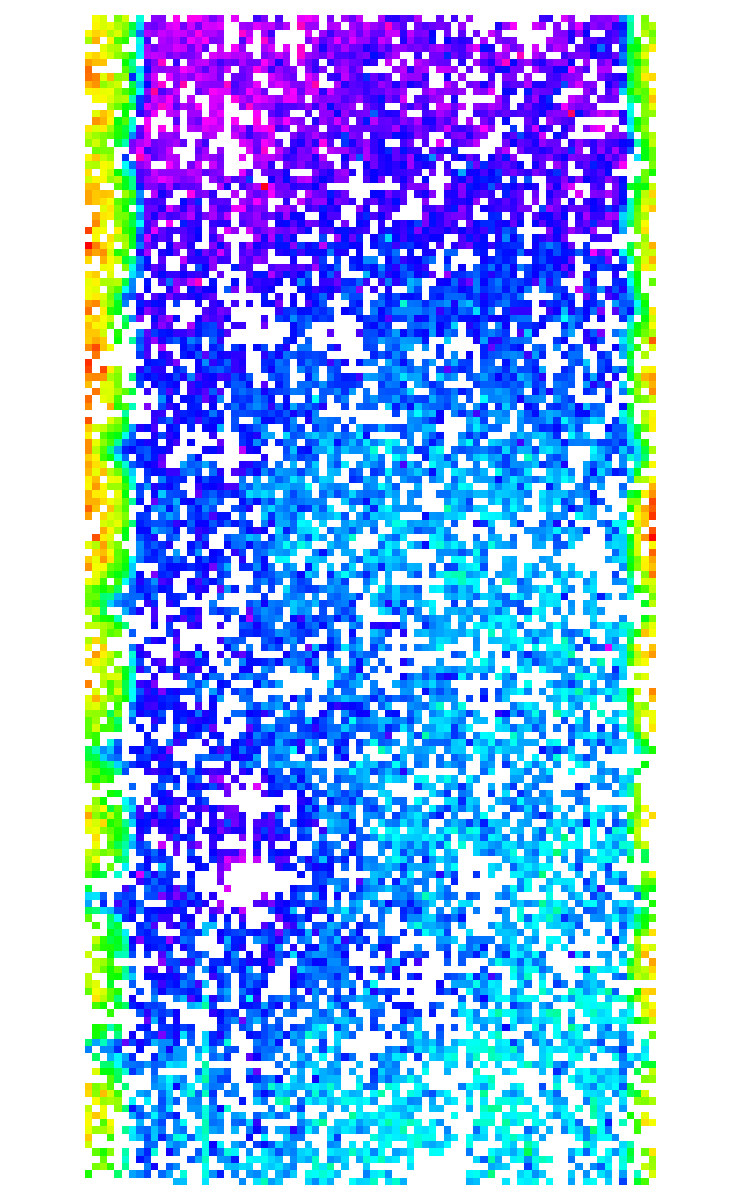}
    \includegraphics[width=0.45\columnwidth]{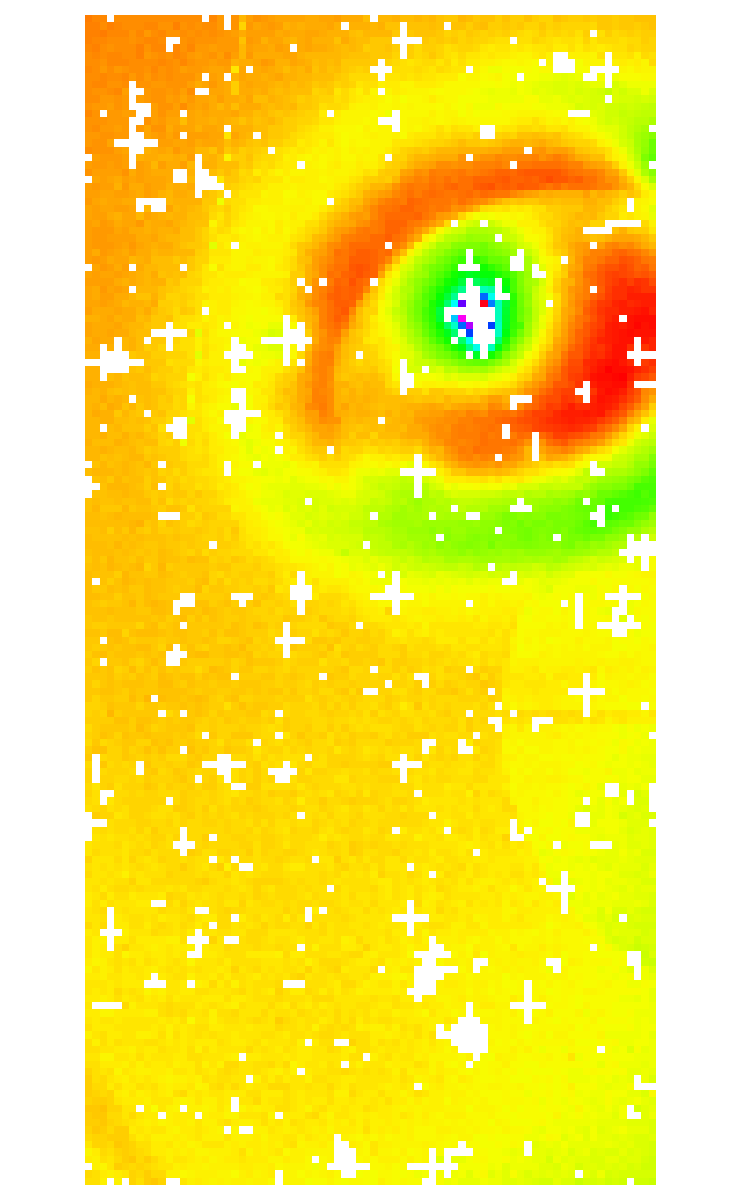}
    }
    \caption{Left: image r418779, CCD\#4. The left and right edge show 5\% difference in background counts. Right: image r531376, CCD\#4. The feature, top right, causes up to 10\% difference relative to the background. The {\sc HaGrid} superpixel map is shown to enhance low level detail. Black areas correspond to rejected superpixels due to high levels of masking. The color scale goes from red (low values) to blue (high values).}
    \label{fig:odd}
\end{figure}

Very occasionally images suffer rare, sometimes unexplained artifacts. Two such examples are shown in figure \ref{fig:odd}. The left hand panel shows reduced counts near the left and right edge of the CCD. This effect is visible in CCDs 2 and 4, but not in CCDs 1 and 3.

The right hand panel of figure \ref{fig:odd} shows the effect of a drop of liquid (water or oil) on the filter. This happened during the two nights of October 10 and 11 2006. The extent and form of the feature changed over time during these nights.

%\newpage
%\begin{appendix} %First appendix
%\onecolumn
\section{Header parameters}
\label{app:header}

The table below provides an example of the header content associated with each CCD image file.  Much of this information is also captured in the metadata table available for download in compressed form as igapsimages.org/data/images/igaps-images.fits.gz.

\onecolumn
\begin{center}
\begin{longtable}[h!]{lll}
\hline
Parameter&Example value&Explanation\\
\hline
BITPIX  &                   8 & Data type of original image                    \\
NAXIS   &                    2 & Dimension of original image                    \\
NAXIS1  &                 2048 & Length of original image axis                  \\
NAXIS2  &                 4096 & Length of original image axis                  \\
PCOUNT  &                    0 & Size of special data area                      \\
GCOUNT  &                    1 & One data group (required keyword)              \\
RUN     &               677729 & Run number                                     \\
OBSERVAT& LAPALMA            & Name of observatory (IRAF style)               \\
%OBJECT  & intphas$_$5023 r     & Title of observation                           \\
OBJECT  & intphas\_5023 r     & Title of observation \\
LATITUDE&            28.761907 & Telescope latitude  (degrees)     \\
LONGITUD&           -17.877559 & Telescope longitude (degrees)     \\
HEIGHT  &                 2348 & [m] Height above sea level.         \\           
SLATEL  & LPO2.5            & Telescope name known to SLALIB      \\           
TELESCOP& INT               & 2.5m Isaac Newton Telescope         \\           
MJD-OBS &        55015.0452356 & Modified Julian Date at start of observation   \\
JD      &      2455015.5452356 & Julian Date at start of observation            \\
PLATESCA&             6.856013 & [d/m] Platescale ( 24.68arcsec/mm)             \\
TELFOCUS&             0.043825 & Telescope focus (metres)                       \\
AIRMASS &             1.021475 & Effective mean airmass                         \\
TEMPTUBE&            13.508864 & Truss Temperature (degrees Celsius)            \\
INSTRUME& WFC              & INT wide-field camera is in use.               \\
WFFPOS  &                    5 & Position-number of deployed filter             \\
WFFBAND & r                  & Waveband of filter                             \\
WFFID   & 214               & Unique identifier of filter                    \\
SECPPIX &                0.333 & Arcseconds per pixel                           \\
DETECTOR& WFC               & Formal name of camera                          \\
CCDSPEED& FAST             & Readout speed                                  \\
CCDXBIN &                    1 & Binning factor in x axis                       \\
CCDYBIN &                    1 & Binning factor in y axis                       \\
CCDSUM  & 1 1                & Binning factors (IRAF style)                   \\
CCDTEMP &              151.959 & [K] Cryostat temperature                       \\
NWINDOWS&                    0 & Number of readout windows                      \\
DATE-OBS& 2009-07-03T01:05:10.8 & Start time of the exposure [UTC]            \\
INHERIT &                    T & Extension inherits primary HDU.                \\
EXTNAME & extension1         & Extension name                                 \\
EXTVER  &                    1 & Extension version number                       \\
IMAGEID &                    1 & Image identification                           \\
DASCHAN &                    1 & Number of readout channel                      \\
WINNO   &                    0 & Number of readout window                       \\
CHIPNAME& A5506-4            & Name of detector chip.                         \\
CCDNAME & A5506-4            & Name of detector chip.                         \\
CCDCHIP & A5506-4            & Name of detector chip.                         \\
CCDTYPE & EEV42-80           & Type of detector chip.                         \\
CCDXPIXE&           0.00001350 & [m] Size of pixels in x.                       \\
CCDYPIXE&           0.00001350 & [m] Size of pixels in y.                       \\
AMPNAME & LH                & Name of output amplifier.                      \\
GAIN    &           2.80000000 & Nominal Photo-electrons per ADU.               \\
READNOIS&           6.40000000 & Nominal Readout noise in electrons.            \\
SATURATE&              64276.0 & Highest value that is unsaturated              \\
%MAXBIAS &       65535.00000000 & Maximum expected bias level                    \\
BIASSEC & [11:50,3:4098]     & Bias pixels.                                   \\
TRIMSEC & [51:2098,3:4098]   & Illuminated pixels.                            \\
RTDATSEC& [2062:4215,13:4212]              & Location in d-space for RTD.     \\
RADESYS & ICRS               & WCS calibrated against Gaia-DR2                \\
EQUINOX &               2000.0 & Equinox of the astrometry                 \\
CTYPE1  & RA \-\-\-ZPN           & Algorithm type for axis 1        \\              
CTYPE2  & DEC\-\-ZPN           & Algorithm type for axis 2        \\              
CRUNIT1 & deg                & Unit of right ascension coordinates            \\
CRUNIT2 & deg                & Unit of declination coordinates                \\
PV2\_1   &                  1.0 & Coefficient for r term                         \\
PV2\_2   &                  0.0 & Coefficient for r**2 term                      \\
PV2\_3   &           213.741679 & Coefficient for r**3 term                      \\
PV2\_5   &                  0.0 & Coefficient for r**5 term                      \\
CRVAL1  &           292.931917 & [deg] Right ascension at the reference pixel   \\
CRVAL2  &           28.6651568 & [deg] Declination at the reference pixel       \\
CRPIX1  &          -329.738223 & [pixel] Reference pixel along axis 1           \\
CRPIX2  &           2945.36999 & [pixel] Reference pixel along axis 2           \\
CD1\_1   &          -1.3972E-06 & Transformation matrix element                  \\
CD1\_2   &          -9.2449E-05 & Transformation matrix element                  \\
CD2\_1   &          -9.2444E-05 & Transformation matrix element                  \\
CD2\_2   &           1.3945E-06 & Transformation matrix element                  \\
STDCRMS &  0.02526049569007405 & Astrometric fit error (arcsec)                 \\
MOONDIST&                 81.0 & Distance to the moon in degrees                \\
MOONALT &    18.29999923706055 & Altitude of the moon above the horizon         \\
MOONPHAS&    83.40000152587891 & Phase of the moon                              \\
SKYLEVEL&               252.99 & Sky level                                      \\
SKYNOISE&    10.96000003814697 & Sky noise                                      \\
PERCORR &               -0.005 & Sky calibration correction (mags)              \\
MAGZPT  &                24.47 & Uncorrected nightly ZP (per second)            \\
MAGZRR  &                 0.02 & Photometric ZP error (mags)                    \\
EXTINCT &                 0.09 & Extinction coefficient (mags)                  \\
PHOTZP  &              28.2187 & mag(Vega) = -2.5*log(pixel value) + PHOTZP     \\
PHOTZPER&                 0.03 & Default 1-sigma PHOTZP uncertainty in IGAPS    \\
PHOTSYS & Vega               & Photometric system                             \\
FLUXCAL & IGAPS-UNIFORM      & Identifies the origin of PHOTZP               \\
SEEING  &             0.753579 & Average FWHM (arcsec)                          \\
ELLIPTIC&   0.1319999992847443 & Average ellipticity                            \\
EXPTIME &                30.07 & [sec] Exposure time adopted                    \\
CONFMAP & iphas\_jul2009&r\_conf.fits Confidence map\\
CHECKSUM& ZfA6ad53VdA3Zd53   & HDU checksum updated 2020-02-11T11:35:56       \\
DATASUM & 1159687462         & data unit checksum updated 2020-02-11T11:35:56 \\
\end{longtable}
\end{center}

%\newpage

%\section{Placement of the $g$ filter mask}
%\onecolumn
%\label{sec:gmask_supp}

%In section~\ref{sec:gband} the impact of a blemish on the $g$ band filter used in the execution of UVEX was described, along with its mitigation.  We show how the mask for flagging affected $g$ magnitudes was applied to the data in figure~\ref{fig:gmask}.  When the $g$ filter was cleaned and replaced in its mount, it did not always go back in oriented as before.  Indeed in the late stages of observation, an effort was made to try to re-orient the filter so that the blemish would fall in front of the cut-out corner of the detector array.  

%\begin{figure*}[h!]\centering
% \resizebox{.85\hsize}{!}{\includegraphics{figures/gmaskplot4.png}}
%\caption{Differences between Pan-STARRS and IGAPS $g$ magnitudes as a function of position in the WFC image plane. Median values are plotted for each 250x250 pixel$^2$ bin. The mask applied for observations made within four different phases of UVEX data collection  are shown.  The diagonal hatched regions represent the placement of the inner $g$-band mask, while the dotted regions indicate the outer mask. Top-left: mask used for observations before June 2006.
%Top-right: mask for observations between June 2006 and December 2013. %Bottom-left: mask for observations between December 2013 and March %2017. Bottom-right: mask for observations after March 2017.}
%\label{fig:gmask}
%\end{figure*}

\newpage
\onecolumn

\end{appendix}
%-------------------------------------------------------------
%                   For appendices and landscape, large table:
%                    in the preamble, use: \usepackage{lscape}
%-------------------------------------------------------------

\end{document}